\documentclass[%
 reprint,
 amsmath,amssymb,
 aps,
floatfix,
]{revtex4-2}

\usepackage{graphicx}
\usepackage{dcolumn}
\usepackage{gensymb}
\usepackage{svg}

\usepackage{bm}
\usepackage{physics}

\usepackage[bookmarks=true,pdftex,colorlinks=true,urlcolor=blue,linkcolor=black,citecolor=blue,breaklinks=true,hypertexnames=false]{hyperref}
\usepackage[capitalize]{cleveref}  

\begin{document}


\title{Hubbard Models for Quasicrystalline Potentials}

\author{E. Gottlob}
 \affiliation{Cavendish Laboratory, University of Cambridge, J. J. Thomson Avenue, Cambridge CB3 0HE, United Kingdom}
\author{U. Schneider}%
\affiliation{%
 Cavendish Laboratory, University of Cambridge, J. J. Thomson Avenue, Cambridge CB3 0HE, United Kingdom}%

\date{\today}

\begin{abstract}
Quasicrystals  are long-range ordered, yet not periodic,
and thereby present a fascinating challenge for condensed matter physics, as one cannot resort to the usual toolbox based on Bloch's theorem. 
Here, we present a numerical method for constructing the Hubbard Hamiltonian of non-periodic potentials without making use of Bloch's theorem and apply it to the case of an eightfold rotationally symmetric 2D optical quasicrystal that was recently realized using cold atoms. We construct maximally localised Wannier functions and use them to extract on-site energies, tunneling amplitudes, and interaction energies. 
In addition, we introduce a configuration-space representation, where sites are ordered in terms of shape and local environment, that leads to a compact description of the infinite-size quasicrystal in which all Hamiltonian parameters can be expressed as smooth functions. This configuration-space picture allows one to efficiently describe the quasicrystal in the thermodynamic limit, and enables new analytic arguments on the topological structure and many-body physics of these models. For instance, we use it to conclude that this quasicrystal will host unit-filling Mott insulators in the thermodynamic limit. 
\end{abstract}

\maketitle
Quasicrystals represent a fascinating middle ground between periodic and disordered materials, they are perfectly long-range ordered without being periodic \citep{senechalQuasicrystalsGeometry1995}.
Quasicrystalline order can naturally arise from an incommensurate projection of a higher-dimensional periodic lattice and thereby enables the investigation of physics of higher dimensions, in particular in the context of topology \citep{langEdgeStatesTopological2012, krausTopologicalStatesAdiabatic2012, krausTopologicalEquivalenceFibonacci2012a, krausFourDimensionalQuantumHall2013a, matsudaTopologicalPropertiesUltracold2014a}, where the resulting structures can inherit topologically protected edge states \citep{krausTopologicalEquivalenceFibonacci2012a, krausTopologicalStatesAdiabatic2012, matsudaTopologicalPropertiesUltracold2014a}. 
Quasicrystals host fractal, self-similar structures both in momentum space \citep{senechalQuasicrystalsGeometry1995} and in their energy spectrum \citep{puigCantorSpectrumQuasiPeriodic2006a}. They also exhibit Anderson localisation \citep{andersonAbsenceDiffusionCertain1958a}, broadly similar to disordered systems.
However, there are crucial differences: in randomly disordered systems in 1D and 2D, the non-interacting spectrum is always fully localised \citep{abrahams_anderson_licciardello_ramakrishnan_1979}. Quasiperiodic systems, on the other hand, can host mobility edges and localisation transitions at finite potential strengths  \citep{aubryAnalyticityBreakingAnderson1980a, diener_transition_2001, biddle_localization_2009,  biddleLocalizationOnedimensionalLattices2011a, ganeshanNearestNeighborTight2015b}.
In the interacting case, localisation can subsist in the form of many-body localisation, whose non-ergodic nature has been the subject of significant attention over the last few years \citep{abaninRecentProgressManybody2017a, vojtaDisorderQuantumManyBody2019, nandkishoreManyBodyLocalizationThermalization2015, altmanUniversalDynamicsRenormalization2015,abanin_colloquium_2019}. There is strong interest in the differences in many-body localisation between quasiperiodic and disordered systems \cite{ khemaniTwoUniversalityClasses2017}, in particular in more than one dimension, where avalanche effects are predicted to destabilize many-body localisation in the latter case \cite{DeRoeck2017}.

To study phase transitions and localisation phenomena, it is convenient to describe the continuum lattice potential as a tight-binding model, i.e.\ as a collection of discrete lattice sites. This tremendously reduces the computational complexity of diagonalising the Hamiltonian, and therefore allows for the study of far larger system sizes.
The key step in constructing a tight-binding Hamiltonian is to generate a set of localised Wannier functions. In periodic lattices, these are constructed as an appropriate superposition of Bloch waves \citep{marzariMaximallyLocalizedWannier2012b}--- which however do not exist for non-periodic potentials. 
For general non-periodic lattices, existing generic methods for calculating Wannier functions are based on imaginary time evolution of trial wave functions \citep{zhouConstructionLocalizedWave2010b}, or rely on full band projections \citep{zhuConstructionMaximallyLocalized2017c}.

Constructing exact tight-binding models for general quasicrystals is difficult because (a) one cannot use Bloch's theorem to construct appropriate Wannier functions and (b) the lack of periodicity typically prevents one from efficiently describing their thermodynamic limit. Several quasiperiodic models, such as Aubry-André models \citep{aubryAnalyticityBreakingAnderson1980a, liu_localization_2015, szaboNonpowerlawUniversalityOnedimensional2018a, ganeshanNearestNeighborTight2015b, biddleLocalizationOnedimensionalLattices2011a, geisslerMobilityEdgeTwodimensional2020},  are explicitly constructed using quasiperiodic perturbations of an initially periodic lattice and thereby inherit the original Wannier functions. These models however represent only particular limits of general quasicrystalline potentials.  

In this paper, we present a method for generating non-periodic Hubbard Hamiltonians without using Bloch's theorem, and apply it to the two-dimensional eightfold rotationally symmetric optical quasicrystal (8QC), see \cref{fig:QClattice}, which has recently been realised with ultracold atoms \citep{viebahnMatterWaveDiffractionQuasicrystalline2019b, sbrosciaObservingLocalization2D2020a}. In addition we introduce a configuration-space description of this quasicrystal, where sites are ordered according to their shape and local environment, which allows to describe the infinite-size quasicrystal in terms of smooth functions in a compact parameter space. This method is similar to configuration-space descriptions employed for stacked bilayer systems \citep{bennettElectricallyTunableStacking2022a, carrRelaxationDomainFormation2018b, massattElectronicDensityStates2017a} and the resulting description directly corresponds to perpendicular spaces widely used in the field of discrete quasicrystals \citep{raiBulkTopologicalSignatures2021a, jagannathanQuasiperiodicHeisenbergAntiferromagnets2012a, ghadimiMeanfieldStudyBoseHubbard2020a, szallasSpinWavesLocal2008a, mirzhalilovPerpendicularSpaceAccounting2020b}.

In \cref{sec:BH}, we present a method for the generation of maximally localised Wannier functions that is applicable to a broad class of quasicrystalline or disordered potentials. We then apply it in \cref{sec:2D8-foldQC} to construct the lowest-band Hubbard Hamiltonian of the 8QC.
In \cref{sec:classification}, we address the description of the quasicrystal in the inifinite-size limit.  We show in \cref{sec:confighubbard} how the 8QC Hubbard Hamiltonian is greatly simplified when re-expressed in configuration space. In \cref{sec:singleband}, we discuss how the validity of the single band picture is impacted by inter-particle interactions. Finally, in \cref{sec:QCMI}, we use the insight gained from the configuration space expression of the 8QC Hubbard Hamiltonian to conclude on the existence of unit-filling Mott insulating phases in the thermodynamic limit.

\section{Tight-Binding Model for non-periodic potentials} \label{sec:BH}

\begin{figure}
  \centering
  \includegraphics[width=\linewidth]{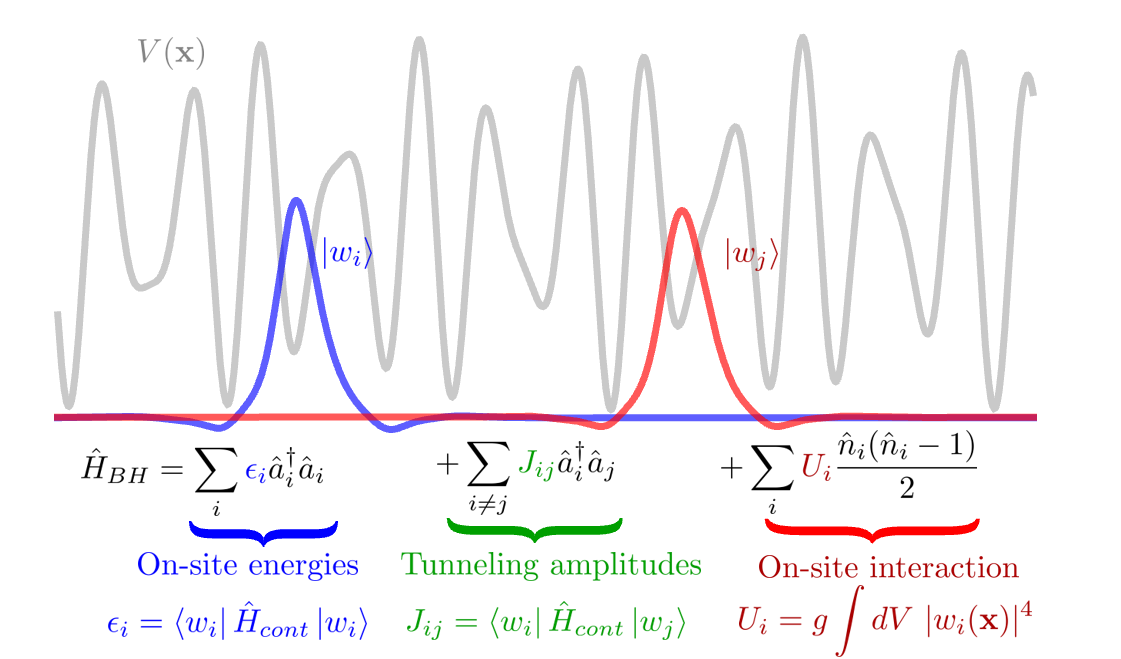}
  \caption{Hubbard models are expressed in terms of on-site energies, tunnelling amplitudes, and on-site interactions, which we compute by constructing maximally localised Wannier functions. In non-periodic systems, all Wannier functions and therefore parameters are site-dependent.}
  \label{fig:NPBH}
\end{figure}
Let us consider a non-periodic lattice described by the \textit{continuum} single-particle Hamiltonian $$\hat{H}_{cont} = \frac{\hat{p}^2}{2m} + \hat{V}(\mathbf{r})\,,$$
where $\hat{p}$ is the momentum operator, $m$ the particle mass and $\hat{V(\mathbf{r})}$ the non-periodic lattice potential.
To generate the tight-binding Hamiltonian, the key step is to obtain an appropriate set of localised single-particle basis states $\left\{\ket{w_{i}} \right\}$ on each site of the lattice, as represented in \cref{fig:NPBH}. We will refer to these states as Wannier functions, even though we are dealing with non-periodic systems where Bloch's theorem does not apply. Our numerical method can be summarised as follows: First, non-orthogonal maximally localized Wannier functions (NOWF) are generated individually on each lattice sites by minimizing the width of a linear combination of eigenstates of the potential -- see below. Second, a Löwdin transformation is applied onto the resulting non-orthogonal set, producing a set of maximally localized and orthogonal Wannier functions (WF). If applied onto a periodic lattice, our method produces the same Wannier functions that would be obtained using the typical Bloch wave formalism.

After constructing the Wannier functions, we generate the Hubbard Hamiltonian (\cref{fig:NPBH}) in the usual way by the explicit evaluation of its matrix elements, namely  on-site energies $\epsilon_i=\bra{w_{i}}\hat{H}_{cont} \ket{w_{i}}$,  hopping amplitudes  $J_{ij} = \bra{w_{i}}\hat{H}_{cont} \ket{w_{j}}$, and on-site interactions $U_{i} = g \int d \mathbf{r} |w_{i}(\mathbf{r})|^4$, where $g = \frac{4 \pi \hbar^2}{m} a $ and $a$ is the scattering length of the considered atomic species. Off-site interactions can also be obtained through the evaluation of integrals involving neighbouring Wannier functions (see \cref{app:offsite})\citep{duttaNonstandardHubbardModels2015}.

\subsection{Maximally localized Wannier Functions in real-space formulation} \label{subsec:NOWF}

Given the non-periodicity of the lattice potential, we cannot rely on Bloch waves for the generation of WFs.
Instead, we start by numerically calculating the single-particle eigenstates $\ket{E_k}$ of the continuum Hamiltonian in a domain of radius $R$ centered around the lattice site at position $\mathbf{r}_{i}$.
We can then express the localized NOWF $\ket{w^{NO}}$ as a linear combination of the single-particle eigenstates within the energy band of interest ($E_{min}\leq E_k\leq E_{max}$):
\begin{equation}
  \ket{w^{NO}}= \sum_{k} c_k \ket{E_k} \,.
  \label{eqwannier}
\end{equation}

We note that,  contrary to periodic crystals, the existence of band gaps separating individual bands is not guaranteed for non-periodic potentials and must be checked individually for each specific lattice potential. 

The coefficients $c_k$ are determined by minimizing the localization criterion \citep{marzariMaximallyLocalizedWannier2012b}
\begin{align} \label{eq:loc}
  \Omega_{i} &\equiv \bra{w^{NO}} (\mathbf{r} - \mathbf{r}_{i})^2 \ket{w^{NO}} \nonumber
\end{align}
subject to the normalisation constraint $\sum_k |c_k|^2 = 1 $.
We can recast this expression as a double sum over all eigenstates $\ket{E_k}$:
\begin{align}
  \Omega_{i} & = \sum_{k,l} c^*_k c_l \bra{E_k} (\mathbf{r} - \mathbf{r}_{i})^2 \ket{E_l} \nonumber \\
      &= \sum_{k,l} c_k^* c_{l} (R^2_i)_{kl} \nonumber \\
      &= \mathbf{c}^\dagger R^2_i \mathbf{c} \,,
\end{align}
where we combine the coefficients $c_k$ into the vector $\mathbf{c}$ and define the hermitian and positive-definite matrix $R^2_i$ with matrix elements $(R_i^2)_{kl} =\bra{E_k} (\mathbf{r} - \mathbf{r}_{i})^2 \ket{E_l} $.

$\ket{w_i^{NO}}$, i.e. the most localized state that can be generated on the lattice site $\mathbf{r}_{i}$, is then directly obtained as the eigenvector of  $(R_i^2)$ with the lowest eigenvalue (which is real-valued thanks to hermiticity). While being maximally localised, the resulting  states $\ket{w_i^{NO}}$ on different sites will not yet be orthogonal.

To obtain an orthogonal set of localised basis states $\ket{w_i}$, the non-orthogonal basis must now be transformed in a way that maintains its localised properties. This is achieved through a Löwdin transformation \citep{lowdinNonorthogonalityProblemWork1970a}
\begin{equation}
  \ket{w_i} = \sum_j S^{-1/2}_{ij} \ket{w_j^{NO}}, \label{eq:Lowdin}
\end{equation}
where $S_{ij} \equiv \braket{w_i^{NO}}{w_j^{NO}}$ is the overlap matrix between NOWFs.
The Löwdin transformation ensures a minimal distance between the orthogonalized and non-orthogonal sets \citep{aikenLowdinOrthogonalization1980a}, i.e.: 
\begin{equation}
  \sum_i \braket{w_i-w_i^{NO}}{w_i-w_i^{NO}} = min
\end{equation}
Therefore, applying a Löwdin transform onto the non-orthogonal maximally localised basis provides us with a maximally localised orthogonal and real-valued basis set. We note that the Löwdin transform fails in case of over-completeness of the non-orthogonal basis, where the overlap matrix does not have maximal rank and therefore cannot be inverted. The success of the Löwdin transform is therefore a good check for over-completeness of the initial non-orthogonal basis set.

In practice, the single-particle eigenstates $\ket{E_k}$ are extracted from a finite-difference formulation of the continuum Schrödinger equation using Lanczos' algorithm \citep{lanczosIterationMethodSolution1950a}, see \cref{appFDS} for details. 
The presented method for constructing WFs becomes exact in the limit of $R \rightarrow \infty$ and vanishing step size for the discretization, but fine grids limit the calculation in practice to relatively modest cut-off radii (typically on the order of 10 lattice sites, i.e. including around 250 to 300 neighbouring lattice sites in the 2D case).
The resulting approximate WFs converge towards the exact WFs when the cut-off radius $R$ becomes much larger than the characteristic size of the WF and we found empirically that implementing the boundary conditions as a hard wall of finite height (cf.\ \cref{fig:boundary} in \cref{app:boundary}) significantly speeds up the convergence, see  \cref{app:conv}. 
We note that the NOWFs on all lattice sites are generated independently of each other; this step  can therefore trivially be parallelised.

\section{Two-dimensional eightfold optical quasicrystal} \label{sec:2D8-foldQC}

\begin{figure}
  \centering
  \includegraphics[width = \linewidth]{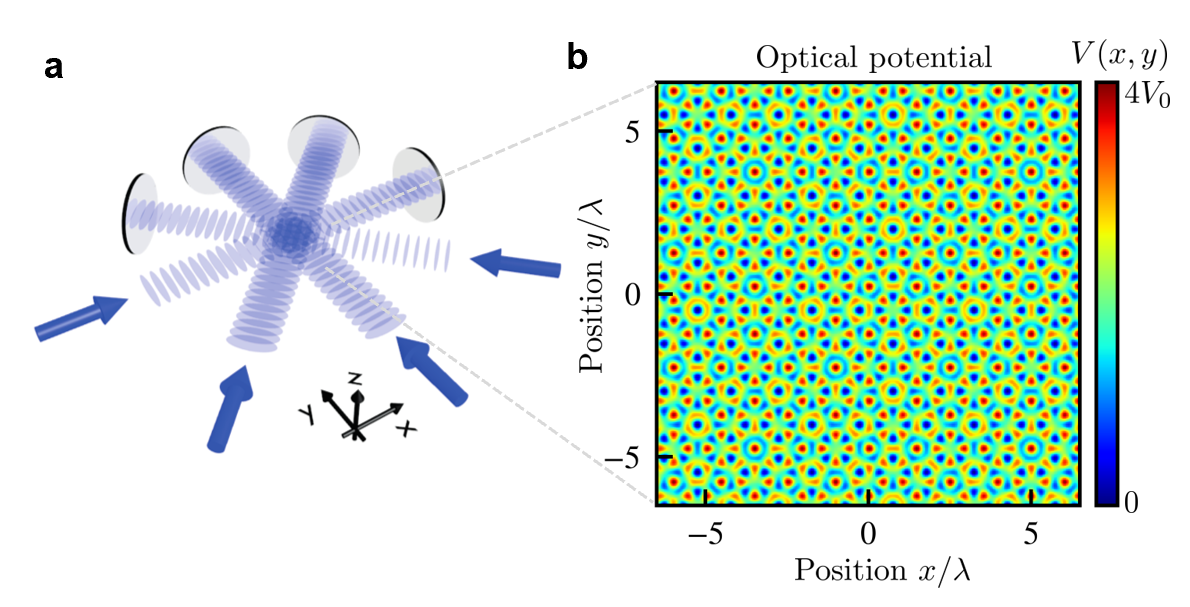}
  \caption{Two-dimensional eight-fold optical quasicrystal. (a) The optical quasicrystal is formed by superimposing two square optical lattice in a single plane with a $45\degree$ angle between them. (b) The resulting optical potential is quasiperiodic.}
  \label{fig:QClattice}
\end{figure}

We now apply the above method to the two-dimensional eightfold quasicrystal (8QC) shown in \cref{fig:QClattice}, which has recently been realised using ultracold atoms \citep{sbrosciaObservingLocalization2D2020a, viebahnMatterWaveDiffractionQuasicrystalline2019b}. 
This continuum quasiperiodic lattice is closely related to the discrete eightfold Ammaan-Beenker lattice \citep{jagannathanEightfoldOpticalQuasicrystal2013a, maceQuantumSimulation2D2016a}. It is formed by superimposing two square optical lattices that are rotated by $45^{\circ}$ with respect to each other and its optical potential (\cref{fig:QClattice}) is given by:
\begin{align}      
     V(\mathbf{r}) &= V_0 \sum_{i=x,y,+,-} \sin^2(\mathbf{k}_i\cdot \mathbf{r}+\phi_i) \nonumber \\
    & \mathbf{k}_i \in \frac{2\pi}{\lambda}\left\{ \begin{pmatrix} 1 \\ 0 \end{pmatrix} \,, \begin{pmatrix} 0 \\ 1 \end{pmatrix} \,, \frac{1}{\sqrt{2}} \begin{pmatrix} 1 \\ 1 \end{pmatrix} \,, \frac{1}{\sqrt{2}} \begin{pmatrix} 1 \\ -1 \end{pmatrix} \right\} \label{eq:QCpotential}
\end{align}
Here, $V_0$ denotes the lattice depths and the  $\mathbf{k}_i$ and $\phi_i$ are the wave vectors and offset phases of the individual lattices created by superimposing laser beams of wavelength $\lambda$.
This potential is clearly long-range ordered, as it is fully deterministic and contains no randomness. At the same time, it cannot be periodic, as 8-fold rotational symmetries are forbidden in periodic lattices \citep{viebahnMatterWaveDiffractionQuasicrystalline2019b}.
In the thermodynamic limit, the physics of the 8QC is independent of the phases $\phi_i$, see \cref{sec:classification}. 

For the remainder of the paper, we will express all energies and lattice depths in units of the \textit{recoil energy} $E_{rec} = \frac{\hbar^2 k^2}{2m}$, and all distances in terms of $\lambda$. For the calculation of $U$, we assume a $20 \, E_{rec}$ deep retro-reflected lattice generated using the same wavelength $\lambda$ along the transverse direction.  While we focus on the Bose Hubbard model, we note that other types of Hubbard models (e.g.\ including longer-range interactions, or describing fermions with spin) can be similarly derived.

\subsection{Extracting Wannier functions}
The generation of WFs for the 8QC presents several challenges. In contrast to e.g.\ the Aubry-Andre model \citep{aubryAnalyticityBreakingAnderson1980a}, the present model cannot be expressed as a perturbation of a periodic model. Therefore, it is a priori not clear whether the lowest part of its single-particle energy spectrum can be described in terms of a single isolated  band. Moreover, even provided such a lowest band exists, it is a priori not clear whether it would correspond to one Wannier function per local minimum. 

\begin{figure}
  \centering
  \includegraphics[width = \linewidth]{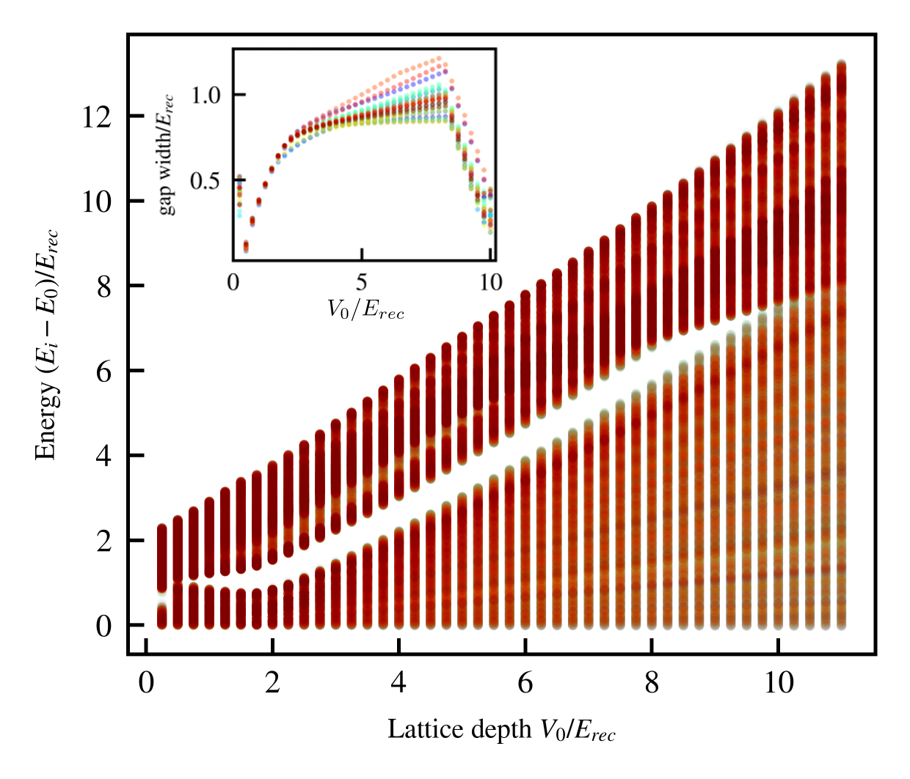}
  \caption{Low-energy single-particle energy spectrum of the 8QC continuum Hamiltonian $H_{cont}$. Different colours represent the  $\approx 750$ lowest bulk eigenstates of 30 different patches of diameter $9 \, \lambda$ that contain around $250$ local minima each. Between $V_0\approx 1$--$10\,E_{rec}$, a clear energy gap separates the  lowest bulk band from higher states. To minimize finite-size effects, boundary conditions are set similarly to \cref{fig:boundary} (b). The inset shows the width of the gap for 30 different patches.}
  \label{fig:gapFDS}
\end{figure}

To investigate whether an isolated lowest band exists, we compute the non-interacting energy spectrum of the bulk of the 8QC by direct numerical diagonalisation of the continuum Hamiltonian, see \cref{fig:gapFDS}. To obtain the bulk energy spectrum, we exclude eigenstates localised on the outer edge of the simulated finite-size patches.
We find that for lattice depths $V_0\approx 1$--$10\,E_{rec}$, the lowest part of the bulk spectrum indeed forms an isolated band that is separated from the rest of the spectrum by a robust gap, independent of the chosen patch. Strikingly, we find that the lowest energy subspace always contains essentially as many states as there are local minima in the finite patch (up to well-understood exceptions, treated in \cref{apprings}). This implies that, for $V_0$ between $1$ to $10 \, E_{rec}$, we can construct a Wannier basis for the lowest band by using one localised Wannier function per local minimum of the potential -- analogous to conventional periodic lattices.
We note that that the 8QC contains less sites per area than a corresponding 2D square lattice, the ratio is equal to the inverse of the silver mean $\frac{2}{1+\sqrt{2}} \approx 0.8284$, see \cref{app:density} for details.

\begin{figure}
  \centering
  \includegraphics[width = \linewidth]{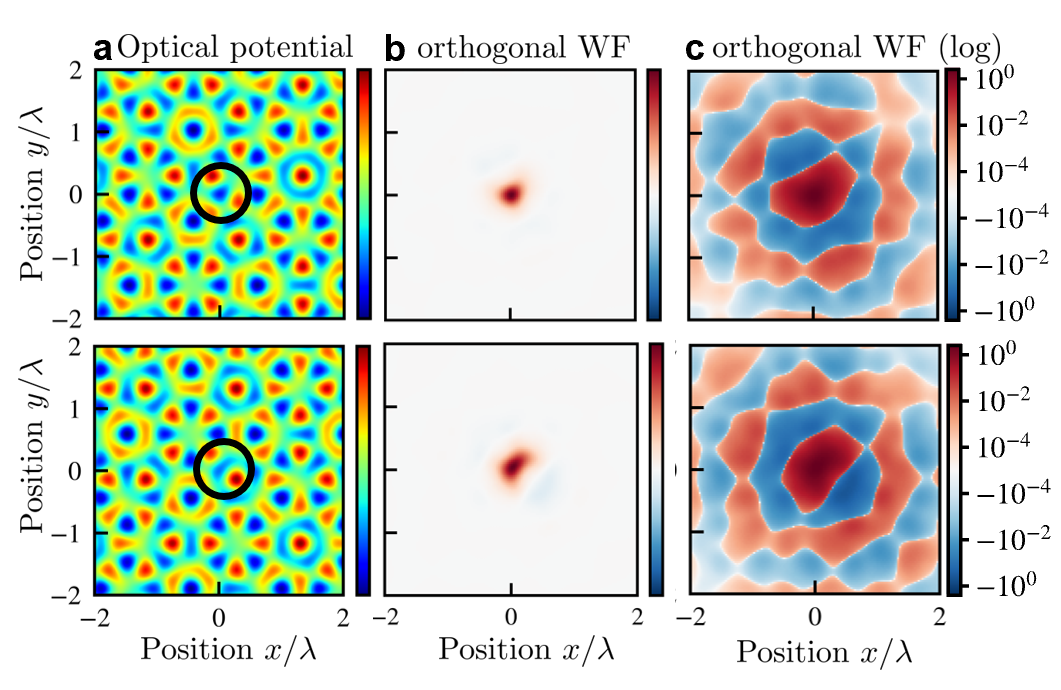}
  \caption{(a): 8QC potential highlighting two different local minima for $V_0 = 2.5 \, E_{rec}$. (b,c): Corresponding orthogonalized Wannier functions plotted on linear and logarithmic scales. As in periodic systems, WFs exhibit oscillating sidelobes with exponential decaying amplitudes  that are clearly visible when plotted on a log scale.
  }
  \label{fig:QCWF}
\end{figure}
We  construct the corresponding  NOWFs by following the method presented in \cref{sec:BH} starting from the eigenstates in the lowest band.
In order to facilitate the convergence of the NOWF already for small cut-off radii $R$, we apply specifically tailored boundary conditions that follow the shape the of the outermost minima; see \cref{app:boundary} for detailed discussion and \cref{app:conv} for numerical convergence checks. Afterwards, we apply a Löwdin transform on the NOWFs to obtain an orthogonal set of maximally localised WFs (\cref{eq:Lowdin}).
\begin{figure}
  \centering
  \includegraphics[width = \linewidth]{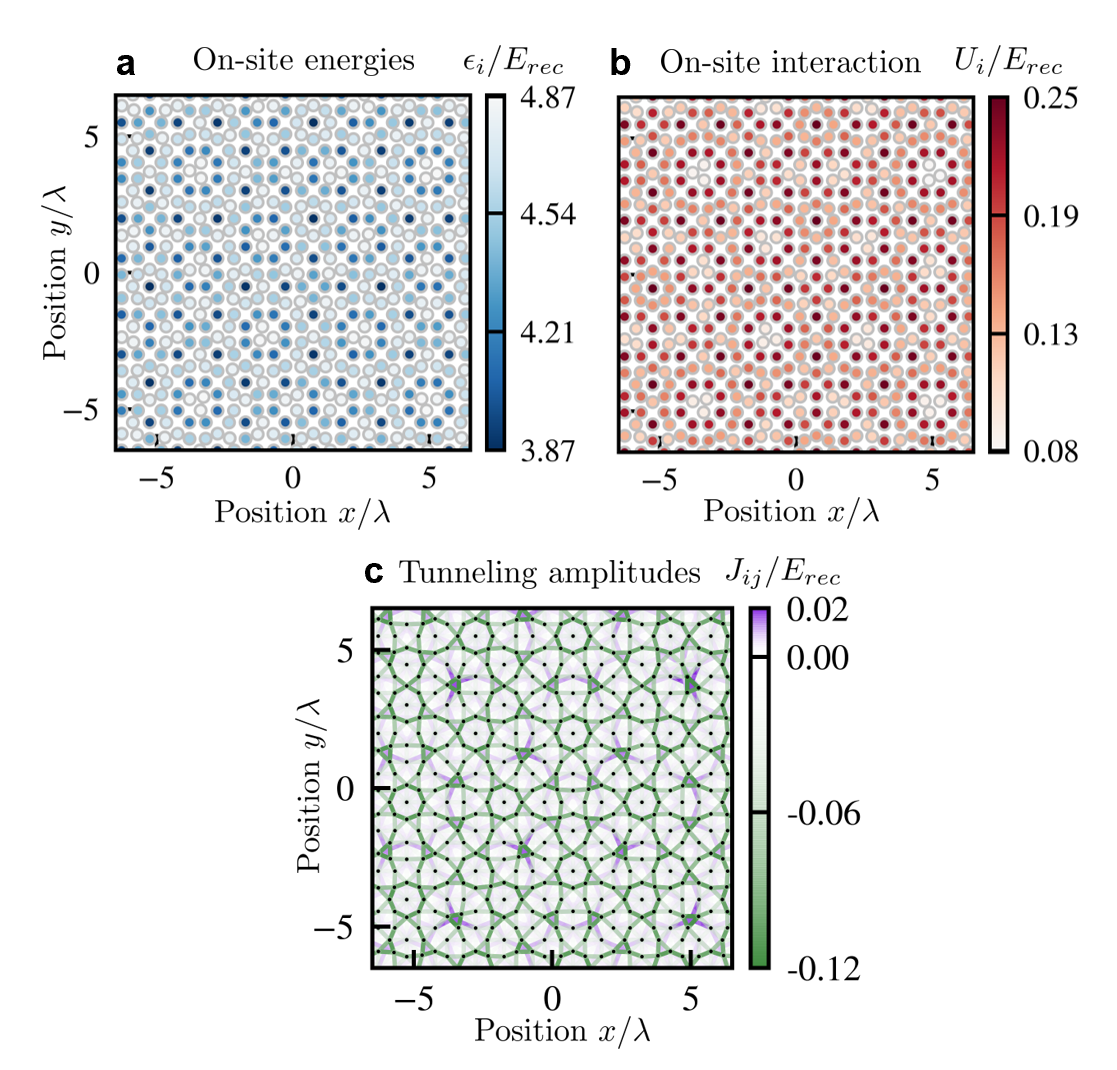}
  \caption{Hubbard parameters  for the 8QC for $V_0 = 2.5\,E_{rec}$ corresponding to the potential shown in \cref{fig:QClattice}. Tunneling amplitudes $J_{ij}$ (both negative and positive) are shown up to 2nd-order neighbours. On-site interactions assume a scattering length of $a = 100\,a_0$ and a $20\, E_{rec}$ transverse lattice.}
  \label{fig:QCBH}
\end{figure}
 Examples of the resulting Wannier functions are shown in \cref{fig:QCWF} and,  similarly to periodic lattices, exhibit exponentially decaying oscillating sidelobes that ensure orthogonality.

\subsection{Bose-Hubbard model}
We next obtain the 8QC Bose-Hubbard Hamiltonian by explicitly computing its matrix elements in the basis of WFs, see \cref{fig:QCBH}, and observe that contrary to simpler models such as Aubry-André models, it is quasiperiodic in all three parameters. Furthermore, on-site energies $\epsilon_i$ and interaction energies $U_i$ are anti-correlated: sites with high $\epsilon_i$ correspond to shallow minima and hence also have a low $U_i$, and vice versa - the same tendency was also noticed in optical lattices with weak quasiperiodic modulation \citep{niederleBosonsTwodimensionalBichromatic2015b}. While the most significant tunneling amplitudes have negative sign, the Hamiltonian also exhibits some small but non-negligible longer range tunneling amplitudes with positive sign. Finally, we show in \cref{app:offsite} that off-site interactions between neighbouring sites can be safely neglected.

\begin{figure}
  \centering
  \includegraphics[width = \linewidth]{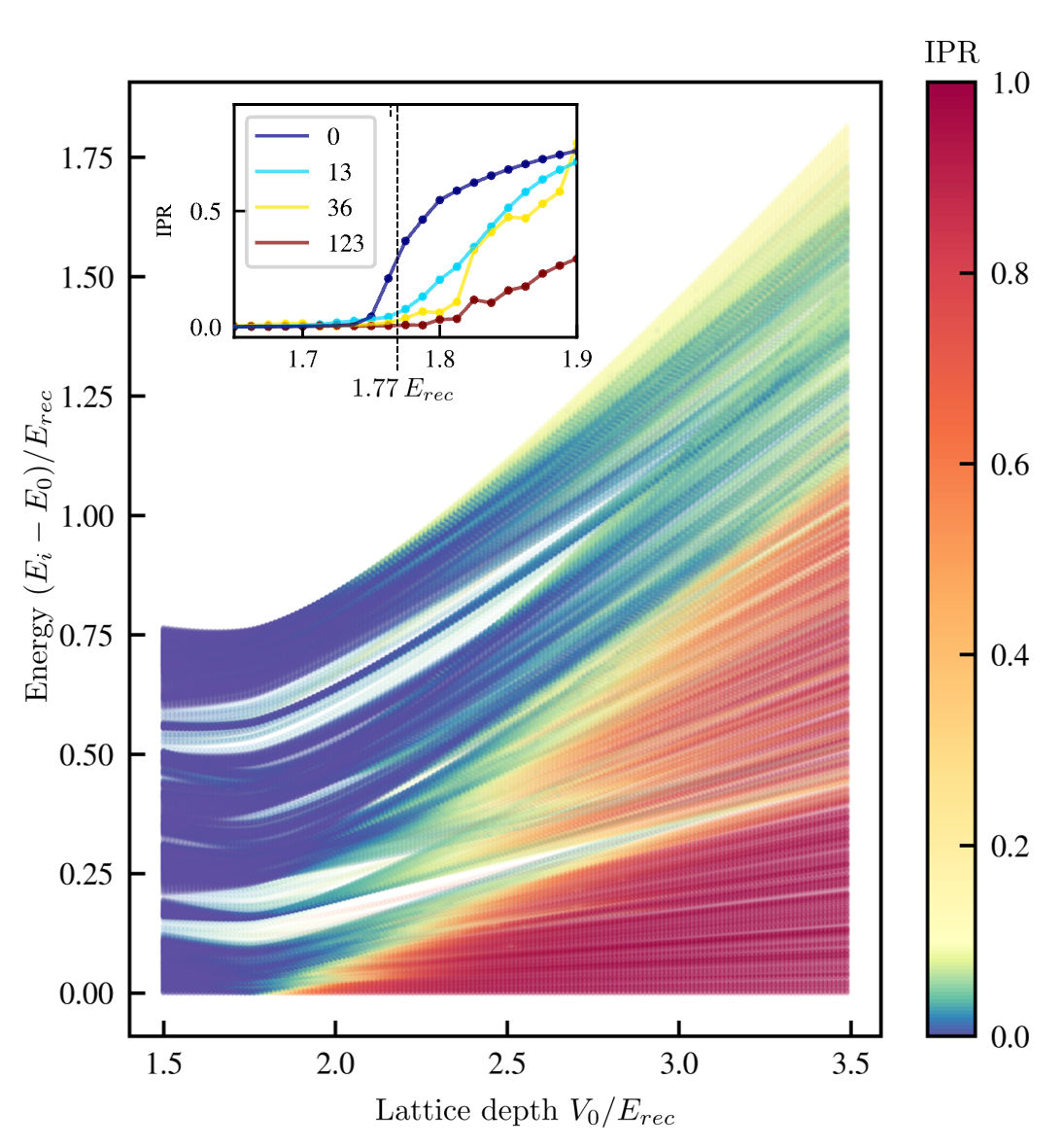}
  \caption{8QC: Non-interacting energy spectrum of the BH Hamiltonian. Color encodes the IPR of the eigenstates. Inset: IPR of the lowest (0) and the 13th, 36th and 136th eigenstates, highlighting the localization transition.}
  \label{fig:IPR}
\end{figure}

\paragraph{\label{sec:singleparticle} Non-interacting energy spectrum}
As an initial benchmark of the resulting Bose-Hubbard model, we use exact diagonalisation in the non-interacting ($a=0$) case to compute the energy spectrum and eigenstates of a finite-sized lattice containing around 2800 sites, i.e.\ ten times more than in the continuum calculation in \cref{fig:gapFDS}. The resulting spectra (\cref{fig:IPR}) contain a series of minigaps at intermediate lattice depths typical for quasiperiodic models. 

\begin{figure}
  \centering
  \includegraphics[width = \linewidth]{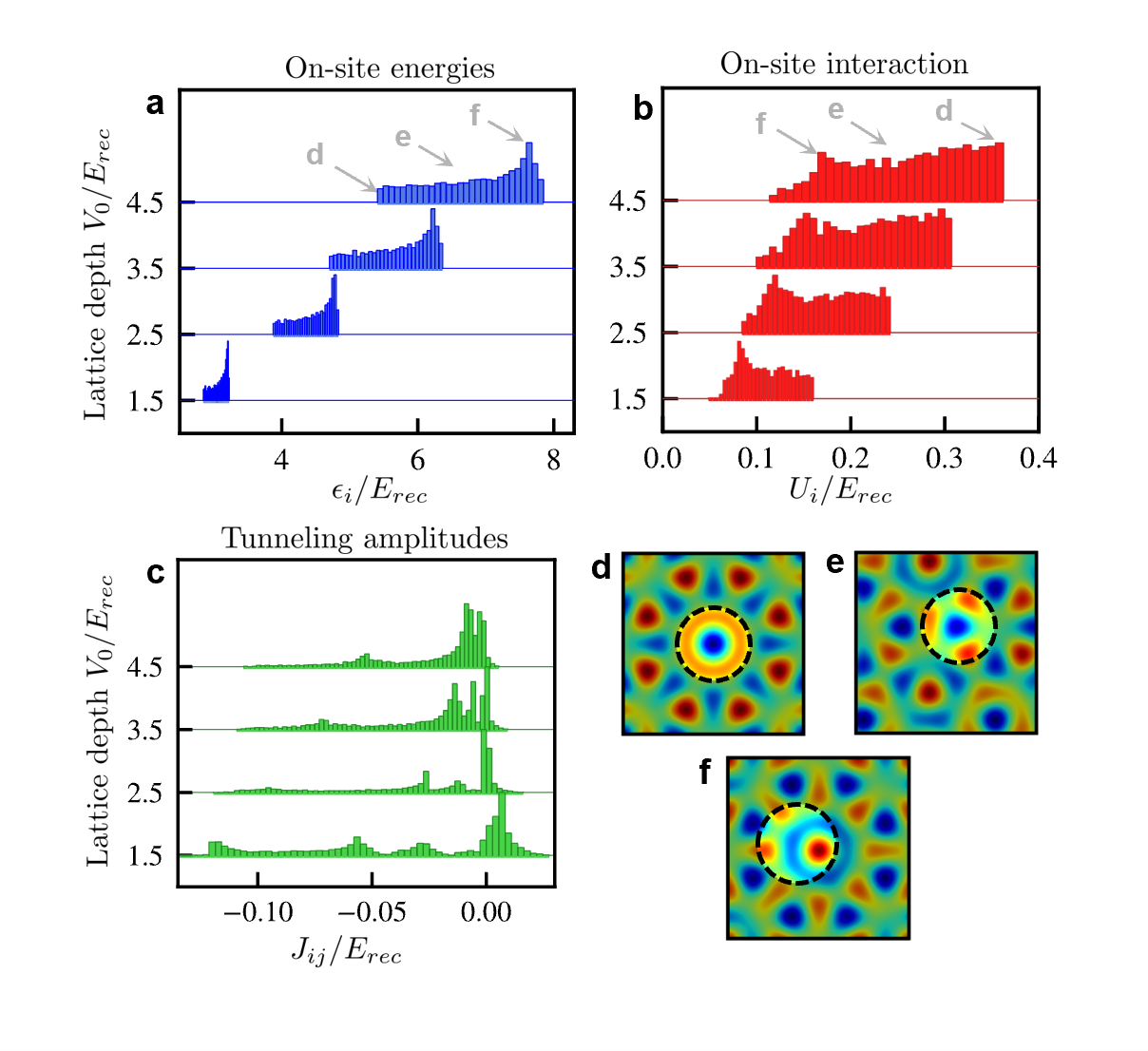}
  \caption{(a,b,c): Histograms of 8QC Hubbard parameters ($\approx 1600$ sites) for various lattice depths $V_0$. Tunneling amplitudes are included up to $2$nd-order neighbours (see \cref{app:neighbours} for definition), and for $|J_{ij}|>10^{-3}\,E_{rec}$. On-site interactions computed for a scattering length $a = 100\,a_0$ and a $20\,E_{rec}$ transverse lattice. 
  (d,e,f): Examples of lattice sites possessing different on-site and interaction energies.}
  \label{fig:histograms}
\end{figure}

The non-interacting physics of the 8QC is governed by the interplay between the tunnelling elements $J_{ij}$ and the energy differences (detunings) $\Delta_{ij}=\epsilon_i-\epsilon_j$ between lattice sites. Resonances, i.e.\ high ratios $J_{ij}/\Delta_{ij}$ between sites favour hybridisation of the corresponding Wannier functions and will lead to delocalisation of the eigenstates.
While all eigenstates are extended for weak lattices, increasing lattice depths $V_0$ lead to decreasing tunneling amplitudes and increasing detunings (cf.~\cref{fig:histograms}). Combined, these two mechanisms strongly decrease the number of resonances and eventually localise all eigenstates.

To quantify the localisation properties of the non-interacting eigenstates $\ket{E_i} = \sum_k c_k^i \ket{w_k}$, we compute their Inverse Participation Ratio (IPR):
\begin{equation}
  IPR_i = \sum |c_k^i|^4 \,.
\end{equation}
An IPR of 1 means that the state is localised on a single lattice site, while the IPR of a fully delocalised state vanishes in an infinitely large system.
The color code in \cref{fig:IPR} represents the IPR of all energy eigenstates as a function of the lattice depth $V_0$. The inset focuses on the IPR of some of the lowest-lying states. It shows that the ground-state undergoes a localisation transition at a critical lattice depth in excellent agreement with the value $V_c = 1.77 \, E_{rec}$ reported in \citep{szaboMixedSpectraPartially2020a,sbrosciaObservingLocalization2D2020a, gautierStronglyInteractingBosons2021b}. Moreover, \cref{fig:IPR} demonstrates that the excited states exhibit a mobility edge separating localised and delocalised states. This is consistent with what is seen in generalised Aubre-Andry Models \citep{ganeshanNearestNeighborTight2015b,biddleLocalizationOnedimensionalLattices2011a, geisslerMobilityEdgeTwodimensional2020}. 

\paragraph{Hubbard parameters\label{sec:QCBH}} 

While  numerical simulations based on the interacting BH Hamiltonian will be left to future work, we can already gain physical insight by inspecting the distributions of on-site energies $p(\epsilon)$, interactions $p(U)$ and tunneling amplitudes $p(J_{ij})$, see \cref{fig:histograms}. The shape of these distributions is very different from what is observed in truly disordered lattice, such as lattices with speckle potentials~\cite{choiExploringManybodyLocalization2016a}. 

\begin{figure}
  \centering
  \includegraphics[width = \linewidth]{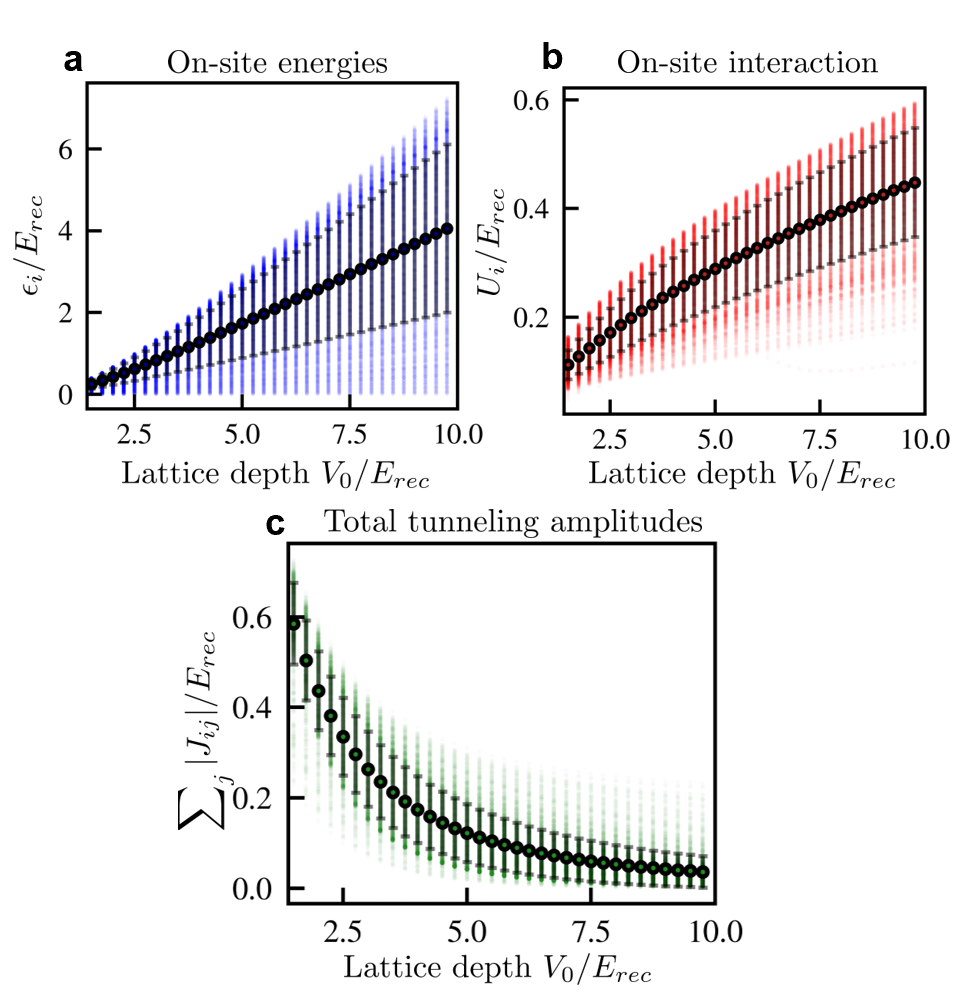}
  \caption{(a,b,c): distribution of 8QC Hubbard parameters ($\approx 1600$ sites) as a function of lattice depth $V_0$. Black dots and bars denote mean values and standard deviations of the distribution. On-site energies are plotted relative to the lowest on-site energy. On-site interactions computed for a scattering length $a = 100\,a_0$ and a $20\,E_{rec}$ transverse lattice.}
  \label{fig:distributions}
\end{figure}

For instance, the distribution of on-site energies contains a sharp maximum reminiscent of a van-Hove singularity (\cref{fig:histograms} a) and increasing the lattice depth $V_0$ causes the width of  $p(\epsilon)$ to increase in an almost linear fashion (\cref{fig:distributions} a). In addition,  the on-site energies form a continuous distribution without any sizeable gaps. 


Increasing lattice depths also leads to increasing interactions strengths, with their mean scaling approximately as $\overline{U}\propto\sqrt{V_0}$, see \cref{fig:histograms} b and \cref{fig:distributions} b, as expected from the decreasing width of the Wannier functions. The upper limit of $p(U)$ closely resembles the value of $U$ expected for a 2D square lattice of depth $2 V_0$. Indeed, the lattice sites sitting at the top of $p(U)$ distribution (\cref{fig:histograms} d) are locally similar to the well of a square lattice of depth $2 V_0$ - both are surrounded by potential barriers of heights close to $4V_0$.
In addition, \cref{fig:histograms} illustrates that irrespective of the lattice depth, the sites located in the deepest potential wells (\cref{fig:histograms} d) are characterised by the lowest on-site energies and highest on-site interaction. Conversely, the highest on-site energies and lowest interaction occur in the shallowest potential wells  (\cref{fig:histograms} f). 

Finally, the average tunneling amplitude (\cref{fig:histograms} c and \cref{fig:distributions} c) decreases broadly exponentially with increasing lattice depth, which is expected from the increasing potential barriers separating lattice sites and the narrower Wannier functions. We also note that a significant share of weak tunneling amplitudes is positive, which we attribute to tunneling between higher-order neighbours, see \cref{app:neighbours} for details.

\section{\label{sec:classification}Configuration space: describing the quasicrystal in the infinite size limit}

There are two main motivations to try to find an alternative description of the quasicrystalline lattice: the absence of periodicity means that there is no simple reciprocal space description, and it would be ideal to have a suitable replacement, i.e.\ a convenient representation on a compact space that is suitable for studying e.g.\ thermodynamic or topological properties of the system. 
In addition, the sites of the 8QC all differ in shape and local surrounding and any finite patch will only contain a subset of all possible sites. 
A priori, one can therefore never be sure whether increasing the simulated system size might introduce additional rare types of lattice sites, changing the results of the simulation. 

To overcome this limitation and arrive at a powerful, compact representation,  we sort the lattices sites based on their shapes and local environments and arrive at a bounded  configuration space that enables us to describe the infinite quasicrystal. This procedure is similar to configuration-space descriptions of stacked bilayer systems \citep{bennettElectricallyTunableStacking2022a, carrRelaxationDomainFormation2018b, massattElectronicDensityStates2017a} and we demonstrate in \cref{app:perp} that it directly corresponds to the perpendicular space of discrete octagonal quasicrystals \citep{raiBulkTopologicalSignatures2021a, jagannathanQuasiperiodicHeisenbergAntiferromagnets2012a, ghadimiMeanfieldStudyBoseHubbard2020a, szallasSpinWavesLocal2008a, mirzhalilovPerpendicularSpaceAccounting2020b}.

\begin{figure}
  \centering
  \includegraphics[width = \linewidth]{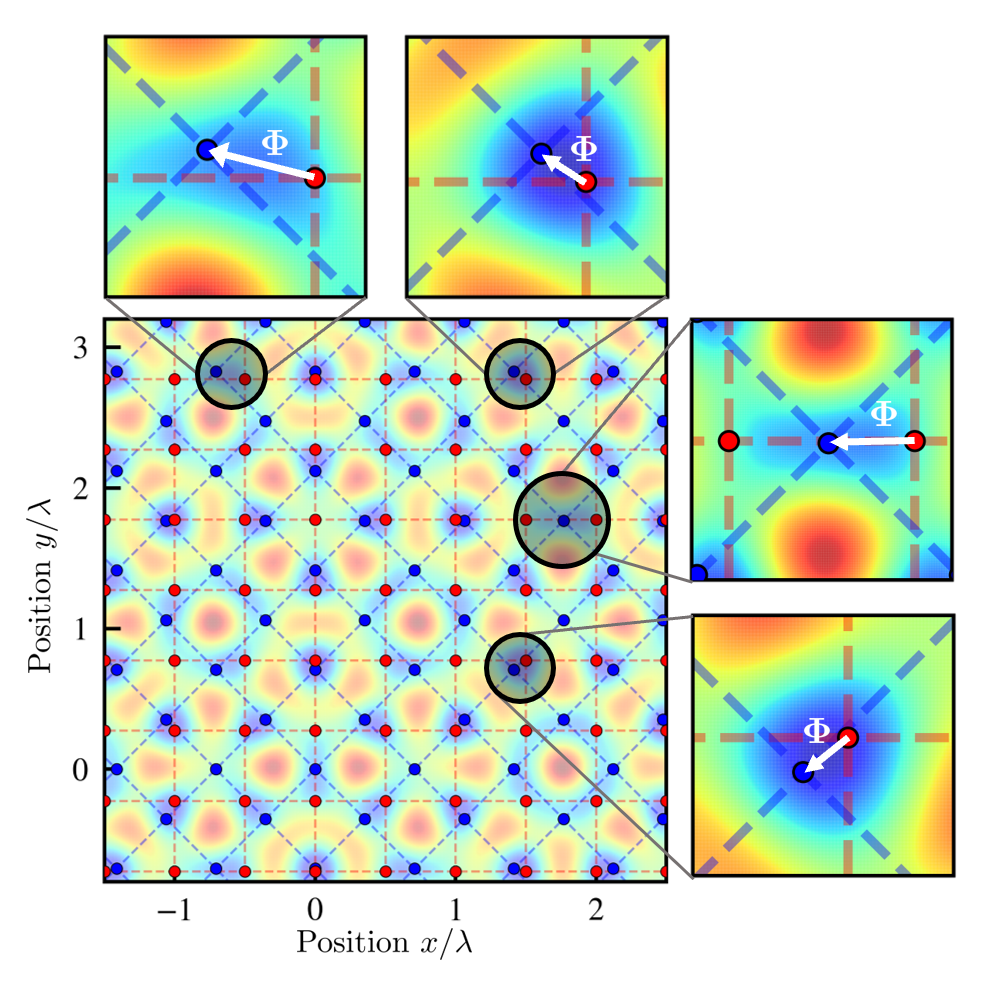}
  \caption{The 8QC potential (background) is formed by superimposing two square lattices (red and blue grids). Insets: For every minimum $\mathbf{r_i}$ of the 8QC, the vector $\mathbf{\Phi}(\mathbf{r_i})$ denotes the displacement between the closest minima of the two square lattices and thereby uniquely defines the local potential. Deep sites correspond to small local displacements $\Phi$, while large $\Phi$ indicate shallow sites.}
  \label{fig:QCgrid}
\end{figure}

We start by recalling that the 8QC is formed by superimposing two square lattices rotated by $45^{\circ}$ (\cref{fig:QClattice}), which is reminiscent of a stacked bilayer system. We refer to the lattice oriented along the $x$ and $y$ axes as the XY lattice. The other square lattice is referred to as the diagonal (D) lattice.
\cref{fig:QCgrid} shows a finite patch of the 8QC potential, where the minima of the XY and D lattices are indicated by red and blue dots. Deep wells in the quasicrystal correspond to closely spaced minima in the XY and D lattices. Conversely, more separated minima of the XY and D lattices result in a shallower minimum in the quasicrystal. Therefore, our mapping procedure characterises each 8QC lattice site in terms of the local displacement $\mathbf{\Phi}$ between the XY and D square lattices.

\begin{figure}
  \centering
  \includegraphics[width = \linewidth]{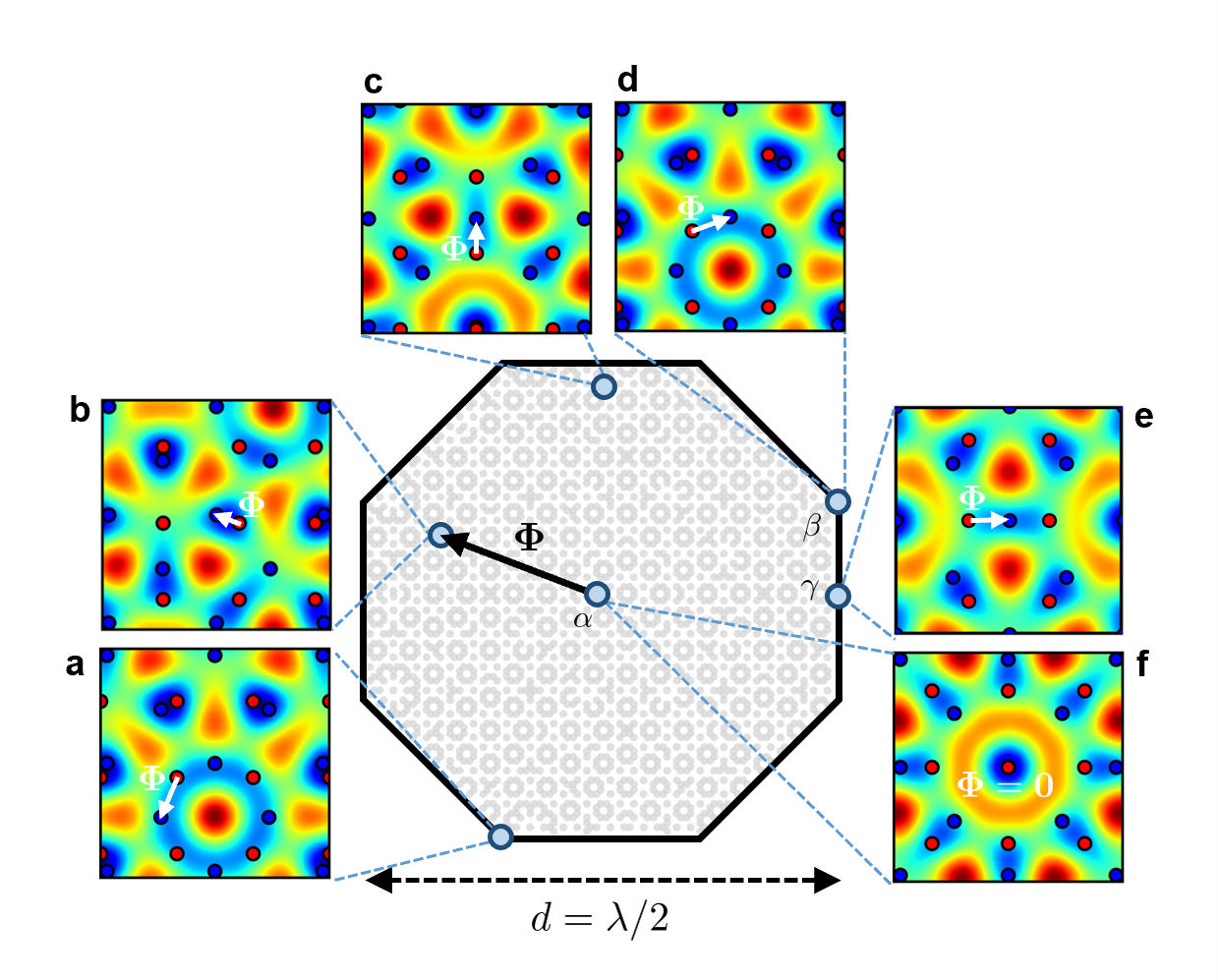}
  \caption{In the infinite-size limit, the configuration-space positions $\Phi$ of the 8QC lattice sites densely and uniformly populate an octagon, whose periodic boundaries are given by the mod operations in \cref{eq:phases_XY}.
  (a,b,c,d,e,f) show the 8QC potential for various positions on the octagon. Dots in the octagon denote the configuration-space positions of 2800 lattice sites.}
  \label{fig:octagon}
\end{figure}

For every minimum in the potential $\mathbf{r_i} = (x_i,y_i)$, we compute its coordinates ($\mathbf{\Phi}_{XY}(\mathbf{r_i})$ and $\mathbf{\Phi}_{D}(\mathbf{r_i})$) within the unit cells  of both the XY  and D lattices:

\begin{align} \label{eq:phases_XY}
  \mathbf{\Phi}_{XY} (\mathbf{r}) &= \left[\left( x+\frac{\phi_1}{k}\right) \mathrm{mod}\; d \right] \, \,  {\mathbf{e}_x} \nonumber\\
              &+ \left[ \left( y+\frac{\phi_2}{k}\right) \mathrm{mod}\; d \right] \, \,   {\mathbf{e}_y}
\end{align}
\begin{align} \label{eq:phases_D}
    \mathbf{\Phi}_{D}(\mathbf{r}) &= \left[\left( \frac{x+y}{\sqrt{2}}+\frac{\phi_3}{k} \right) \mathrm{mod}\; d \right] \, \,  {\mathbf{e}_+} \nonumber \\
                & + \left[\left( \frac{x-y}{\sqrt{2}}+\frac{\phi_4}{k} \right) \mathrm{mod}\; d \right] \, \,  {\mathbf{e}_-} \,
\end{align}
Here, $d=\lambda/2$ denotes the lattice constant of the square lattices, $\mathbf{e}_x$, $\mathbf{e}_y$ are the unit-vectors along the $x$ and $y$ directions,  $\mathbf{e}_\pm = \frac{\mathbf{e}_x\pm \mathbf{e}_y}{\sqrt{2}}$,  and the $\phi_i$ are the four phases introduced in \cref{eq:QCpotential}. 
For every site $\mathbf{r_i}$, the vector $\mathbf{\Phi}(\mathbf{r_i}) \equiv \mathbf{\Phi}_{XY}(\mathbf{r_i}) - \mathbf{\Phi}_{D}(\mathbf{r_i})$ then encodes the local displacement between the XY and D lattices and thereby fully describes the shape of the minimum and its local surroundings.  
As shown in \cref{fig:octagon}, the vectors $\mathbf{\Phi}(\mathbf{r_i})$ describing the  sites of the 8QC form an octagon of inscribed radius $d/2$, where the size stems from the periodicity of the two square lattices.

This configuration-space representation has the following properties: (1) In the infinite-size limit, the octagon is densely and uniformly populated with lattice sites (see \cref{app:neighbours}) and can therefore be used to derive statistical estimates about the lattice. We notice that this is identical to the perpendicular spaces of octagonal discrete  quasiperiodic lattices \cite{mirzhalilovPerpendicularSpaceAccounting2020b, mirzhalilovPerpendicularSpaceAccounting2020b}. (2) Due to the aperiodicity of the lattice, every point $\mathbf{\Phi}$ within the octagon corresponds to one unique lattice site.  
(3) Analogous to the Brillouin zone in periodic crystals, the  $\mathrm{mod}\; d$ operation implies periodic boundary conditions for this configuration space, i.e.\ every edge of the octagon can be identified with the opposing edge. These periodic boundary conditions imply that the octagon possesses the topology of a two-hole torus -- it is an orientable surface with genus 2 \citep{spurrierSemiclassicalDynamicsBerry2018a}. In particular, the 8 corners of the octagon are one unique point. (4)  Symmetry points or lines of the octagon directly correspond to symmetry points or lines of the quasicrystal. For example, the center and corners of the octagon corresponds to the two possible global 8-fold  rotational symmetry centers of the lattice (\cref{fig:octagon} f and a). 
By construction, these two can never be found together in the same realisation of the 8QC, and most choices of $\phi_i$ will lead to none of them. 

This configuration-space construction allows one to draw several conclusions regarding the infinite-size 8QC and will enable novel studies on topology in quasicrystals, as it provides a compact manifold on which e.g.\ Berry curvature and related quantities can be defined.

\subsection{Hubbard model in configuration space} \label{sec:confighubbard}
We can now re-express the BH Hamiltonian of the 8QC entirely in the compact and densely populated configuration space.
This is achieved by mapping the real-space coordinates $\mathbf{r}_i$ of all sites of the 8QC to the corresponding  $\mathbf{\Phi}(\mathbf{r}_i)$:

\begin{align} 
H_{BH} &= \sum_{\mathbf{\Phi}} \epsilon(\mathbf{\Phi})  \hat{a}_{\mathbf{\Phi}}^\dagger \hat{a}_{\mathbf{\Phi}} \nonumber\\
& + \sum_{\mathbf{\Phi} \neq \mathbf{\Phi}'} J(\mathbf{\Phi}, \mathbf{\Phi}') \hat{a}_{\mathbf{\Phi}}^\dagger \hat{a}_{\mathbf{\Phi'}} \nonumber \\
& + \sum_{\mathbf{\Phi}} \frac{U(\mathbf{\Phi})}{2} \hat{n}_{\mathbf{\Phi}}(\hat{n}_{\mathbf{\Phi}}-1) 
\label{eq:BHconfig} \end{align}

\begin{figure}
  \centering
  \includegraphics[width =1\linewidth]{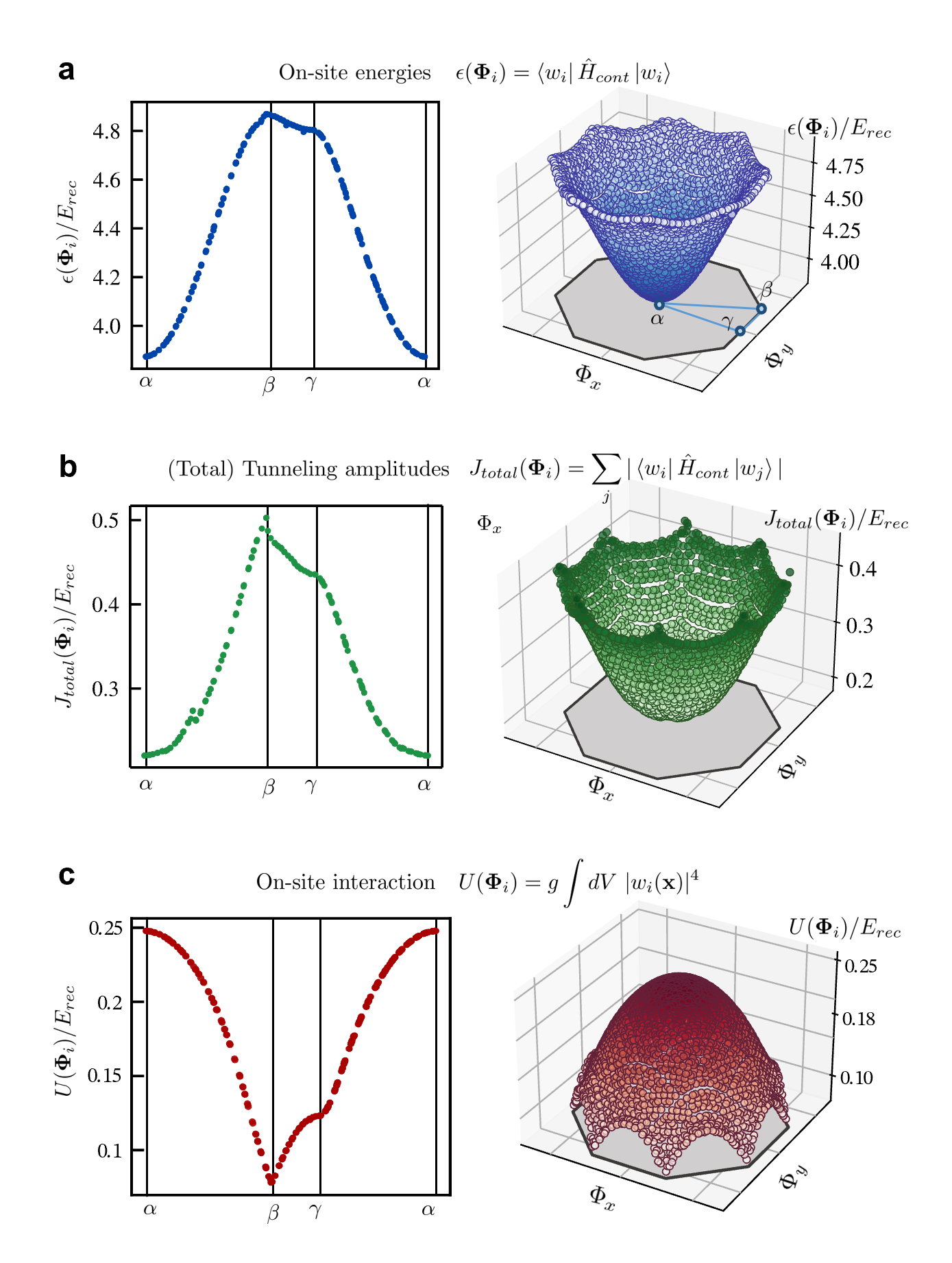}
  \caption{8QC: The Hubbard Hamiltonian can be re-expressed in configuration space, i.e.\ in the space of local displacements $\mathbf{\Phi}$. In this representation, the Hubbard parameters form smooth and 8-fold symmetric surfaces $\epsilon(\mathbf{\Phi})$, $U(\mathbf{\Phi})$ and $J(\mathbf{\Phi}, \mathbf{\Phi}')$. The $\alpha, \beta$ and $\gamma$ points are shown in (a). Figures show $2800$ sites. On-site interactions are computed for a scattering length $a = 100\,a_0$ and $20\, E_{rec}$ transverse lattice. For clarity, (b) shows the total tunneling amplitude per site $J_{total}(\mathbf{\Phi_i}) = \sum_{i\neq j} |J_{ij}|$. The individual tunneling amplitudes $J(\mathbf{\Phi}, \mathbf{\Phi}')$ in configuration space are discussed in \cref{app:neighbours}.}
  \label{fig:BHoctagon}
\end{figure}
Here, $\hat{a}_\Phi$ is the annihilation operator for the WF at the site with coordinate $\Phi$ in configuration space.
This expression emphasises that in configuration space -- which is a compact and uniformly dense space with periodic boundaries -- the 8QC is entirely described by the functions $\epsilon(\mathbf{\Phi}), J(\mathbf{\Phi}, \mathbf{\Phi}'), U(\mathbf{\Phi})$. These are shown in \cref{fig:BHoctagon} and reveal a striking property: contrary to the fractal structure in real space, the Hubbard parameters form smooth functions in configuration space. This directly follows from the construction of configuration space, where an infinitesimal move in $\mathbf{\Phi}$ implies an infinitesimal relative displacement between the XY and D square lattices and hence  a smooth change of the resulting potential.
In turn, the Wannier function hosted on the corresponding lattice site will also undergo an  infinitesimal change, resulting in the observed smooth changes of all local properties. As a consequence, on-site energy, interaction,  tunneling amplitudes, or any other local property must be smooth in configuration space. 

Two important consequences follow from the smoothness of Hubbard parameters in configuration space. First, arbitrary large Hubbard Hamiltonians can now be obtained 
at negligible computational cost, for example by computing Wannier functions for a finite number of points in configuration space and interpolating between them, or by direct interpolation of the Hubbard parameters. Second, the physics of the quasicrystal is, for sufficiently large system sizes, unaffected by the specific values of the phases $\phi_i$. While they amount to global translations in configuration space, the dense sampling ensures that \cref{eq:BHconfig} remains effectively unaffected.

Turning to the shape of these surfaces, we observe that the sites close to the centre of the octagon (\cref{fig:octagon} f) correspond to almost perfectly overlapping minima of both square latices and are hence located within the deepest potential wells in real-space. Therefore, they possess low on-site energy $\epsilon(\mathbf{\Phi})$, high on-site interaction energy $U(\mathbf{\Phi})$, and low total tunneling amplitudes $J_{total}(\mathbf{\Phi_i}) = \sum_{i\neq j} |J_{ij}|$. Conversely, sites corresponding to the corners of the octagon (\cref{fig:octagon} a,f) are located on shallow and high-lying potential wells. They possess the highest on-site energies, the lowest on-site interaction and highest tunneling amplitudes. 

As a sidenote, we notice that the on-site energies (\cref{fig:BHoctagon} a) can be approximated by a simple analytical expression: 
\begin{equation} \label{eq:onsite}
  \epsilon (\mathbf{\Phi}) \approx \Delta_0 + \Delta \sum_{i=1}^4 \sin^2 \left( \frac{\mathbf{k}_i}{|\mathbf{k}_i|}\cdot \mathbf{\Phi} \right).
\end{equation}
for $V_0$ between $1.5$ to $10 \, E_{rec}$. This approximation, whose form is surprisingly reminiscent of the lattice potential in $\cref{eq:QCpotential}$, has an average relative root-mean-square error smaller than $1 \%$.
The individual tunneling amplitudes $J(\mathbf{\Phi}, \mathbf{\Phi}')$ are more intricate and are discussed in \cref{app:neighbours}, where we also show how the configuration space picture allows us to unambiguously define a hierarchy of first-, second-, and higher order neighbours in a matter reminiscent of the fractal structure found in momentum space \cite{viebahnMatterWaveDiffractionQuasicrystalline2019b}. First-order neighbours and the lines connecting them  form the well-known Ammaan-Beenker tiling. We emphasize, however, that there can be significant tunneling elements also connecting 2nd-order neighbours. In contrast to the Ammaan-Beenker tiling, the 8QC is hence not bipartite. In future studies, it will be of interest to determine whether exact closed-form solutions can be obtained for the functions $\epsilon(\mathbf{\Phi})$, $U(\mathbf{\Phi})$ and $J(\mathbf{\Phi}, \mathbf{\Phi}')$.

\subsection{Validity of single-band picture}
\label{sec:singleband}

\begin{figure}
  \centering
  \includegraphics[width = \linewidth]{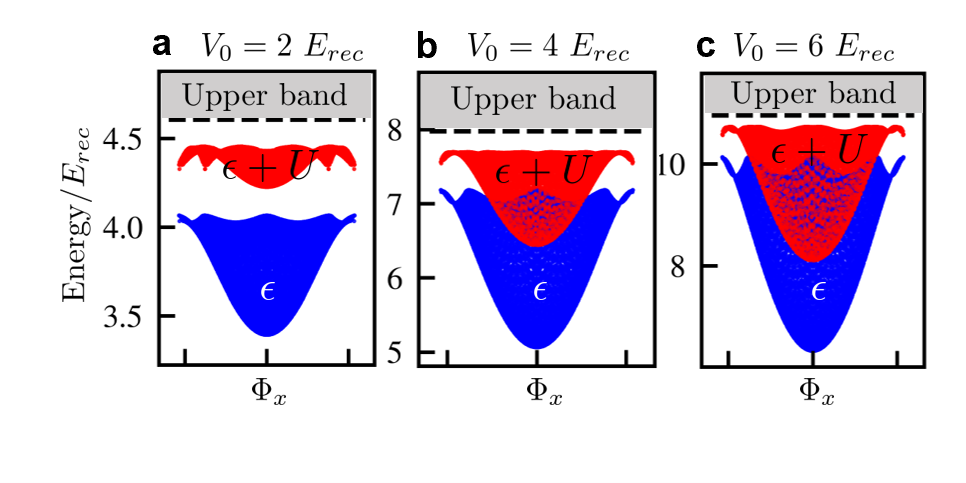}
  \caption{8QC: Estimation of maximal interaction energy achievable without exciting atoms into the higher bands using configuration space. The blue spectrum indicates the on-site energies $\epsilon(\mathbf{\Phi})$, while the red dots correspond to the energy needed for a second particle on the same site, i.e., $\epsilon(\mathbf{\Phi}) + U(\mathbf{\Phi})$. We consider $20\,E_{rec}$ transverse confining lattice and a scattering length $a = 400\,a_0$.}
  \label{fig:QCMI}
\end{figure}
The BH Hamiltonian presented in this work only considers the lowest band, i.e., one WF per lattice site. This is sufficient as long as temperature, chemical potential, and on-site interaction energies are smaller than the energy difference to the first excited WFs. The extension to more WFs per lattice site is left to future research, but we can extract the approximate onset of the second band already from the spectra of the continuum Hamiltonian for small systems, see \cref{fig:gapFDS}. These are indicated by the black dashed line in \cref{fig:QCMI}, where the blue dots indicate the on-site energies $\epsilon(\mathbf{\Phi})$, and the red dots indicate the energy needed for a second particle on the same site, i.e.,\ $\epsilon(\mathbf{\Phi})+U(\mathbf{\Phi})$.
We consider a transverse confining lattice of depth $V_z = 20 \, E_{rec}$ and find that for lattice depths $V_0$ between $1$ and $9 \, E_{rec}$, we can reach a scattering length $a_{max} \approx 400 \, a_0$ before doubly occupied sites begin to overlap with higher bands (\cref{fig:QCMI}). This indicates the validity of the singe-band model in the relevant regimes.

\subsection{Mott-insulating phases \label{sec:QCMI}} 
The precise ground state phase diagram of the interacting 8QC will require large scale numerical simulations (e.g.\ Quantum Monte-Carlo \citep{polletReviewMonteCarlo2013}) that will be facilitated by the Hubbard model developed in this work. One might, in analogy to other disordered or quasiperiodic lattices, expect it to contain superfluid and Mott-insulating phases that are separated by the compressible Bose-glass phase~\citep{giamarchi_schulz_1988, fisher_boson_1989, kiskerBoseglassMottinsulatorPhase1997a,
hitchcockBoseglassSuperfluidTransition2006a,
senguptaQuantumGlassPhases2007a,  
rouxQuasiperiodicBoseHubbardModel2008, 
polletAbsenceDirectSuperfluid2009a,
soylerPhaseDiagramCommensurate2011a,
niederleBosonsTwodimensionalBichromatic2015b, 
johnstoneMeanfieldPhasesUltracold2019b, johnstone_mean-field_2021, gautierStronglyInteractingBosons2021b,ciardiFinitetemperaturePhasesTrapped2022}. While the existence of Mott-insulating phases is a priori not clear, we can already draw some conclusions from the developed Hubbard model:

In the atomic limit, where tunneling can be neglected, and in the presence of a band gap above the lowest band, an incompressible Mott-insulating (MI) phase with one atom per site (i.e.\ $\frac{2}{1+\sqrt{2}} \approx 0.83$ atoms per $(\lambda/2)^2$, see \cref{app:density}) will exist whenever
\begin{equation}
\epsilon(\mathbf{\Phi}) + U(\mathbf{\Phi})> \underset{\mathbf{\Phi'}}{\text{max}}\left(\epsilon(\mathbf{\Phi'})\right) \quad\forall\mathbf{\Phi},
\label{eq:MI}
\end{equation}
i.e., whenever the interaction dominates over the spread in on-site energies, see \cref{fig:QCMI}.
This suggests that for relatively large lattice depths (but with $V_0 < 10\,E_{rec}$ such that a finite band gap exists) and sufficiently high scattering lengths, there will always be an incompressible MI phase with unit filling. We note, however, that excitations of this Mott insulator or its extension to finite temperatures would likely not be accurately described by the current BH model, as double occupancies could hybridize with the excited band.

Furthermore, \cref{fig:distributions} highlights that with increasing lattice depth, the spread in $\epsilon(\mathbf{\Phi})$ grows faster than the average on-site interaction. \cref{eq:MI} therefore implies that the transition from Bose glass to MI will for deeper lattices shift to larger scattering lengths. This is in stark contrast to a periodic lattice, where the transition from superfluid to MI shifts to smaller scattering lengths for deeper lattices.

As a third important conclusion, the continuous distribution of on-site energies $\epsilon(\mathbf{\Phi})$ in the thermodynamic limit directly implies that, at least in the atomic limit, there are no incommensurate Mott phases below unit filling, as such states would always be gapless and compressible. This suggests that the MI states with fractional fillings found in recent quantum Monte-Carlo simulations of the continuum model \cite{gautierStronglyInteractingBosons2021b} might be limited to finite system sizes, where configuration space by necessity is only sampled coarsely.

\section{Conclusion}

We presented a general numerical method for computing the Wannier functions and Hubbard Hamiltonians of non-periodic potentials. This method was then applied to construct the Bose-Hubbard Hamiltonian of the two-dimensional eightfold symmetric optical quasicrystal (8QC). 
As a benchmark, we reproduced the localisation transition in the non-interacting ground state and obtained excellent agreement with earlier results. 

In a second part, we introduced a configuration-space representation of the 8QC. This representation, inspired by existing schemes for incommensurate bilayer systems, enables the description of the quasicrystal in the infinite-size limit by ordering the lattice sites in terms of their shape and local surrounding.
We showed that the Hubbard model of an infinite 8QC can be re-expressed on a dense and compact octagon with periodic boundary conditions. In this representation, the Hubbard parameters take the form of smooth functions.

This Hubbard model opens the door to large-scale numerical simulations of quasicrystalline optical lattices, and the developed configuration space  enables new analytic arguments about the many-body physics and topological structure of these models. 

In future studies, it will be of interest to apply the configuration-space picture to other quasiperiodic lattices, such as Aubry-André models and models interpolating between quasicrystalline and Aubry-André limits.

\begin{acknowledgments}
This work was partly funded by the European Commission ERC Starting Grant
QUASICRYSTAL, the EPSRC Grant EP/R044627/1 and Programme Grant DesOEQ
(EP/P009565/1). We are thankful to Shaurya Bhave, Georgia Nixon, Lee Reeve, Bo Song, and Jr-Chiun Yu from the Quasicrystal Team, along with Callum Duncan, Andrew Daley, Arjun Ashoka, Jonathan Crabbe, and Arta Safari for helpful discussions.

\end{acknowledgments}

\appendix

\section{Finite-Difference Schrödinger equation} \label{appFDS}

We use the Finite-Difference Schrödinger Equation (FDS) for the numerical solution of the Schrödinger eigenvalue equation $\hat{H}\ket{\psi_i} = E_i \ket{\psi_i}$, illustrated below in the one-dimensional case for simplicity.
Deriving the FDS Hamiltonian consists of writing the matrix elements of the Hamiltonian $\hat{H} = -\frac{1}{4 \pi^2}\Delta + V(x)$ (written in units of $E_{rec}$) in a discretised position basis $\ket{x_i}$ with grid spacing $\delta x=\frac{L}{N}$, where $L$ is the system size and $N$ the number of grid points.

Using a finite-difference approximation, we can write the Laplacian as $$ \Delta \psi(x) \approx \frac{\psi(x+\delta x)-2\psi(x)+\psi(x-\delta x)}{(\delta x)^2 }\,.$$ Therefore, its matrix elements in the discretised basis are $$\bra{x_i} \Delta \ket{x_j} = \frac{\delta_{i+1,j}-2\delta_{i,j}+\delta_{i-1,j}}{(\delta x)^2}\,.$$ The potential operator $\hat{V}$ is diagonal in the discretised basis and its matrix elements are $\bra{x_i}\hat{V}\ket{x_j} = V(x_i) \ \delta_{i,j}$. Consequently, we can write the matrix elements of the FDS Hamiltonian as

\begin{align}
  \bra{x_i}H\ket{x_j} & = -\frac{1}{4\pi^2}\bra{x_i} \Delta \ket{x_j}
   + \bra{x_i}\hat{V}\ket{x_j} \\
   & = -\frac{1}{4\pi^2} \frac{\delta_{i+1,j}
  -2\delta_{i,j}+\delta_{i-1,j}}{(\delta x)^2}
   + V(x_i) \ \delta_{i,j} \,.
\end{align}

We use a numerical matrix eigenvalue solver based on Lancsoz' algorithm \citep{lanczosIterationMethodSolution1950a} to obtain the lowest eigenvalues $E_i$ and corresponding eigenvectors $\ket{E_i} = \sum c_j^i \ket{x_j}$ of the finite-difference Hamiltonian $H$ with open boundary conditions. 

The FDS algorithm is naturally limited by Nyquist's theorem. In order for the algorithm to be accurate, the inverse of the discretisation step should always be at least twice the maximal momentum contained in the Fourier transform of the considered state. Consequently, for a given discretisation step size, the precision of the obtained solution decreases for higher-lying states.

\section{Boundary conditions for generating the Wannier functions}\label{app:boundary}

\begin{figure}
    \centering
    \includegraphics[width = \linewidth]{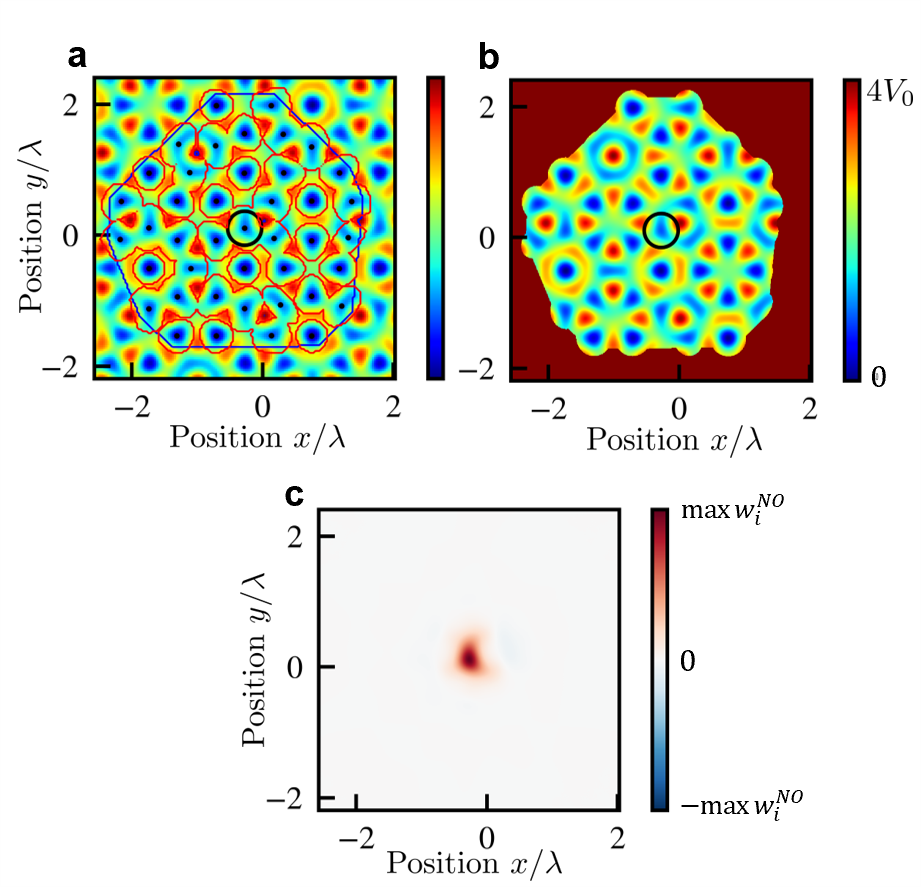}
    \caption{(a): The boundary conditions for generating the NOWF (b) are generated by combining two criteria. One is the convex hull of the set of all sites within the cut-off radius enlarged by $0.17 \, \lambda$ (blue line). The second combines individual boundaries around each lattice site (red), that consist of the contour lines sitting $15\%$ above the bottom of the potential wells, enlarged by a factor of 2.5. The resulting boundary wall is shown in b. Afterwards, a non-orthogonal Wannier function (c) is constructed by localising a linear combination of eigenstates around the central minimum (dark circle).}
    \label{fig:boundary}
\end{figure}

The boundary conditions for calculating the NOWFs consist of a hard wall of height $4 \, V_0$, whose shape is generated in two steps. We first compute the convex hull of the set of all lattice sites within the cut-off radius $R$, and then enlarge it by $0.17 \, \lambda$ (blue line on \cref{fig:boundary}). This results in a boundary that still strongly affects the wavefunction on the wells closest to the cut-off radius. Therefore, we calculate a second boundary that closely matches the shapes of the wells (red line on \cref{fig:boundary}). This is created from the contour line sitting $15\%$ of the total amplitude of the optical potential (which is $4 V_0$) above the bottom of the potential wells, and enlarged by a factor of 2.5. Joining these two then leads to the boundary condition shown in \cref{fig:boundary} b that improves convergence for small cut-off radii.

\section{Convergence checks} \label{app:conv}

The convergence of the BH  parameters is controlled by two parameters: the grid spacing $\delta x$ and the cut-off radius $R$ for the generation of NOWFs.

\begin{figure}
  \centering
  \includegraphics[width = \linewidth]{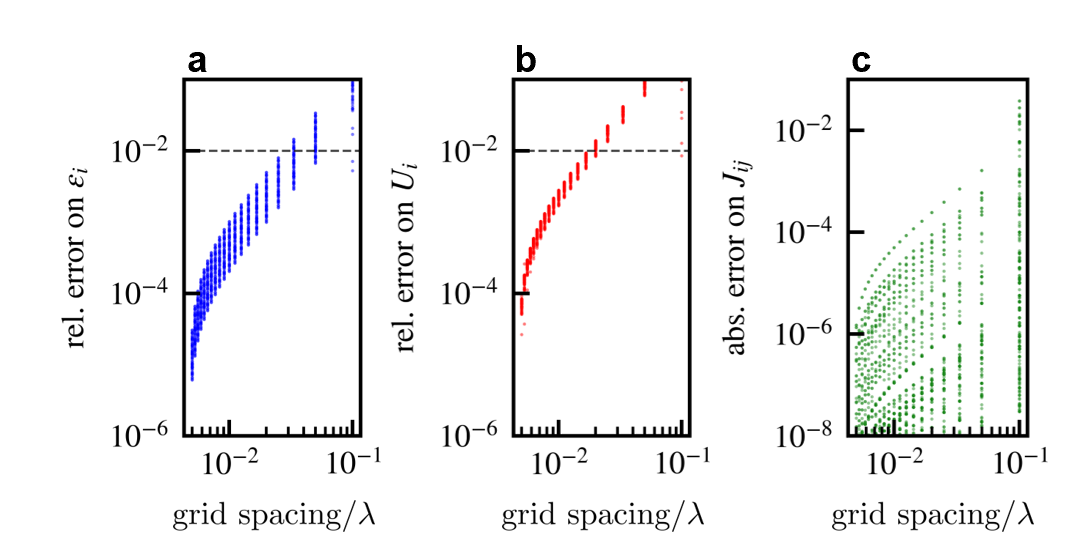}
  \caption{Convergence of on-site energies (a), on-site interaction (b) and nearest-neighbour tunneling amplitudes up to 2nd-order neighbours (c) as a function of the numerical grid spacing $\delta x$ at $V_0 = 9\,E_{rec}$. System containing $63$ lattice sites. Errors are obtained by comparing the results with a "converged" solution computed for a grid spacing of $\delta x_{min}=4.76 \times 10^{-3} \, \lambda$.}
  \label{fig:stepconv}
\end{figure}

\cref{fig:stepconv} shows the result of a convergence study in the grid spacing at a lattice depth of $9 \, E_{rec}$. We estimate the convergence by comparing the Hubbard parameters (on-site energies, interaction and nearest-neighbour tunneling amplitudes) to a "converged" solution computed with the smallest grid spacing of $\delta x_{min}=4.76\times 10^{-3} \, \lambda$ on a system of 63 lattice sites. In the case of the tunneling elements, we considered the absolute error instead of the relative error, as very small tunneling amplitudes can have large relative errors without affecting the physics of the model. 
As a result, we set the grid spacing for all lattice depths to $\delta x=0.03\,\lambda$, resulting in an accuracy of $\approx1\%$.
This requirement in grid spacing is most stringent at the highest lattice depth, as deeper lattices reduce the spread of the WFs.

\begin{figure}
  \includegraphics[width =\linewidth]{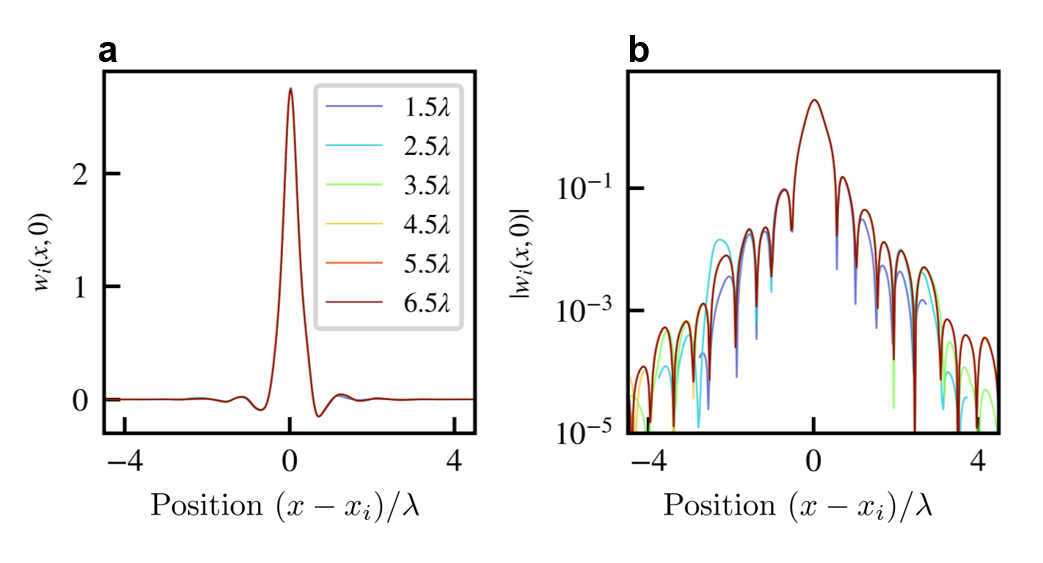}
    \centering
  \caption{Convergence of an 8QC Wannier function as a function of cut-off radius $R$ on linear (a) and logarithmic (b) scales. Figure only shows the horizontal cross section of the 2D Wannier function. $V_0 = 1.5 \, E_{rec}$.}
  \label{fig:conv_WF}
\end{figure}

\cref{fig:conv_WF} illustrates the convergence of a Wannier function for increasing cut-off radii $R$, for a relatively shallow depth of $V_0 = 1.5 \, E_{rec}$. As seen on a logarithmic scale, the relevant sidelobes (and thereby the tunneling elements) quickly converge when $R$ is increased. 

\begin{figure}
  \includegraphics[width =\linewidth]{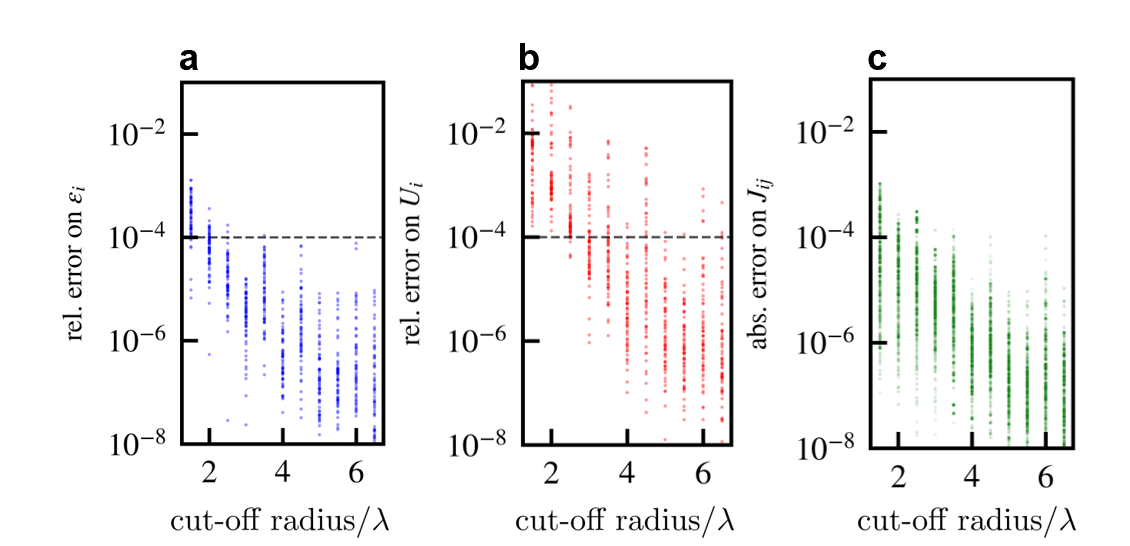}
  \centering
  \caption{Convergence of the on-site energies(a), on-site interaction (b) and tunneling amplitudes up to 2nd-order neighbours (c) as a function of the cut-off radius $R$. $V_0 = 1.5\,E_{rec}$. System containing $66$ lattice sites. Errors are obtained by comparing the results with a "converged" solution computed for a cut-off radius of $7 \, \lambda$.}
  \label{fig:cut-offconv}
\end{figure}

To assess the convergence of the WF more quantitatively, \cref{fig:cut-offconv} shows the effect of varying the cut-off radius $R$ on the convergence of on-site energies, interaction and nearest-neighbour tunneling amplitudes of $66$ lattice sites for $V_0 = 1.5 \, E_{rec}$. We estimate convergence by comparing them with the result of a "converged" solution computed with a cut-off radius of $7 \, \lambda$. 
We expect the requirement in cut-off radius $R$ to be more stringent at low lattice depth where the Wannier functions are more spread out. As a result, we set the cut-off radius to $R=4 \, \lambda$ for all lattice depths (except where stated otherwise in the text). 

As an additional test, we used the same $\delta x$ and $R$ to generate the Hubbard Hamiltonian of a finite-size square periodic lattice, and compared the nearest-neighbour tunnelling amplitudes to the result expected from maximally localised Wannier functions computed using Bloch waves. In the range $V_0 = 2$ to $10\,E_{rec}$, the relative error in on-site interaction and nearest-neighbour tunneling  was always below $3\times 10^{-2}$. 

\section{Exponential localisation of the Wannier functions}\label{app:decay}

\begin{figure}
  \centering
  \includegraphics[width = \linewidth]{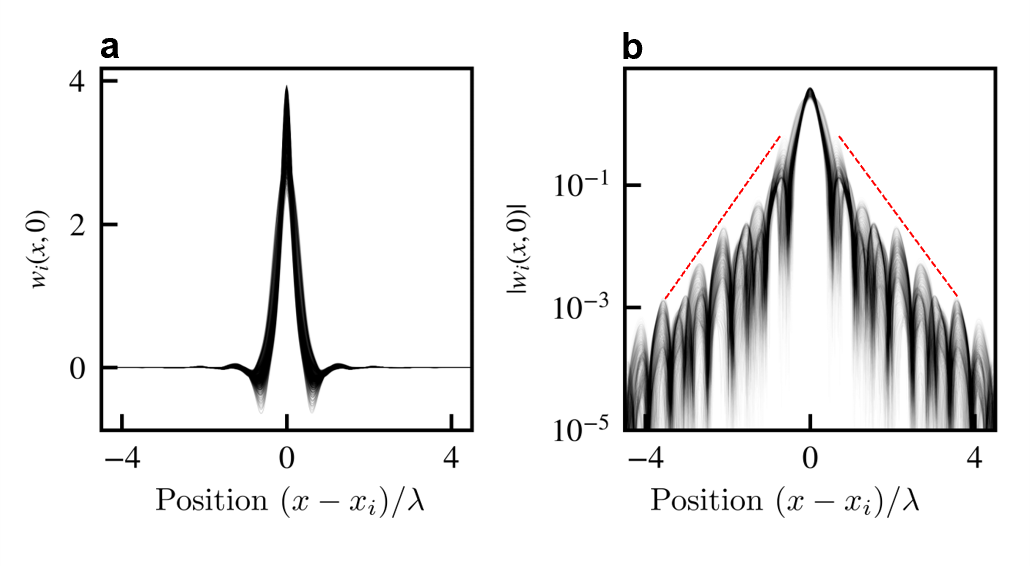}
  \caption{Cross-section of 1600 different Wannier functions on linear (a) and logarithmic (b) scales. $V_0 = 1.5 \, E_{rec}$. Red lines indicate the exponentially decaying sidelobes.}
  \label{fig:wf_decay}
\end{figure}

\cref{fig:wf_decay} shows cross-sections of 1600 different 8QC Wannier functions, obtained for a low lattice depth of $V_0=1.5 \,E_{rec}$, i.e.,\ below the ground state localisation transition. These exhibit exponentially decaying sidelobes that are clearly visible as a linear decay in logarithmic scale (red lines). 

\section{Off-site interactions} \label{app:offsite}

\begin{figure}
    \centering
    \includegraphics[width = \linewidth]{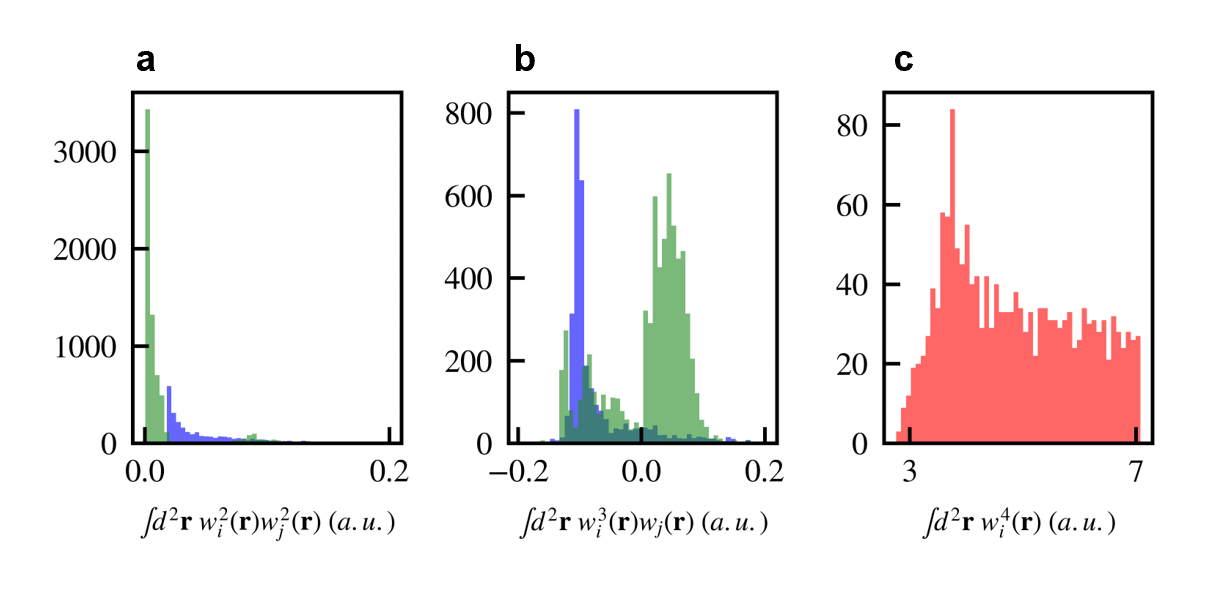}
    \caption{(a,b): Histograms of two-body amplitudes between neighbouring sites. 1st-order neighbours in blue and 2nd-order neighbours in green. (c):  Corresponding histogram for two-body on-site interaction. Lattice depth is $V_0 = 1.5\,E_{rec}$.}
    \label{fig:offsite}
\end{figure}

Atoms on neighbouring sites can in principle interact through various two-body processes due to the overlap of the corresponding Wannier functions \citep{mazzarella_giampaolo_illuminati_2006, sowinski_dutta_hauke_tagliacozzo_lewenstein_2012,duttaNonstandardHubbardModels2015}. The matrix elements of these processes involve integrals of the form: 
\begin{equation}
    U_{ijkl} \propto \int d^2 \mathbf{r} w_i^*(\mathbf{r)}w_j^*(\mathbf{r)}w_k(\mathbf{r)}w_l(\mathbf{r)} \,,
\end{equation}
which considers Wannier functions located on different lattice sites. The most significant processes involve just one pair of sites ($U_{iijj}$ and $U_{iiij}$).
To check whether these off-site processes could be significant for the 8QC, we explicitly compute the overlap integrals for pairs of neighbouring sites, in a shallow lattice ($V_0=1.5 \, E_{rec}$) containing around $1600$ sites. As seen in \cref{fig:offsite}, these off-site processes are always at least one order of magnitude smaller than the lowest on-site interaction energies. We can therefore safely neglect them for all lattice depths above $1.5 \, E_{rec}$, where the off-site processes are even further suppressed due to the increased confinement of the WFs.

\section{Density of sites of the eight-fold optical quasicrystal}\label{app:density}
The configuration-space picture can be employed to derive the exact density of lattice sites in the eight-fold optical quasicrystal: we first  note that in the limit of vanishingly weak beams in the diagonal $k_+$ and $k_-$ directions, the resulting potential contains as many sites as the usual square lattice -- which has a density $n_{square}$ of 1 site per $(\lambda/2)^2$. In addition, the configuration space of this lattice with weak diagonal beams now constitutes a densely populated square of side $d$. Increasing the weak diagonal beams to restore the eight-fold symmetry adds new periodic boundary conditions along the diagonal directions in configuration space and reduce the square to the octagon shown in \cref{fig:octagon}. Since the density in configuration space remains constant, we can directly infer that the ratio of the density of sites in the 8QC ($n_{8QC}$) to the square lattice is given by the ratio of the area of the octagon to the square, i.e., the inverse of the silver mean: 

\begin{equation}
    \frac{n_{8QC}}{n_{square}} = \frac{2}{1+\sqrt{2}} \approx 0.8284
\end{equation} 

\section{Exceptional minima at the boundary of configuration space} \label{apprings}
\begin{figure}
  \centering
  \includegraphics[width = \linewidth]{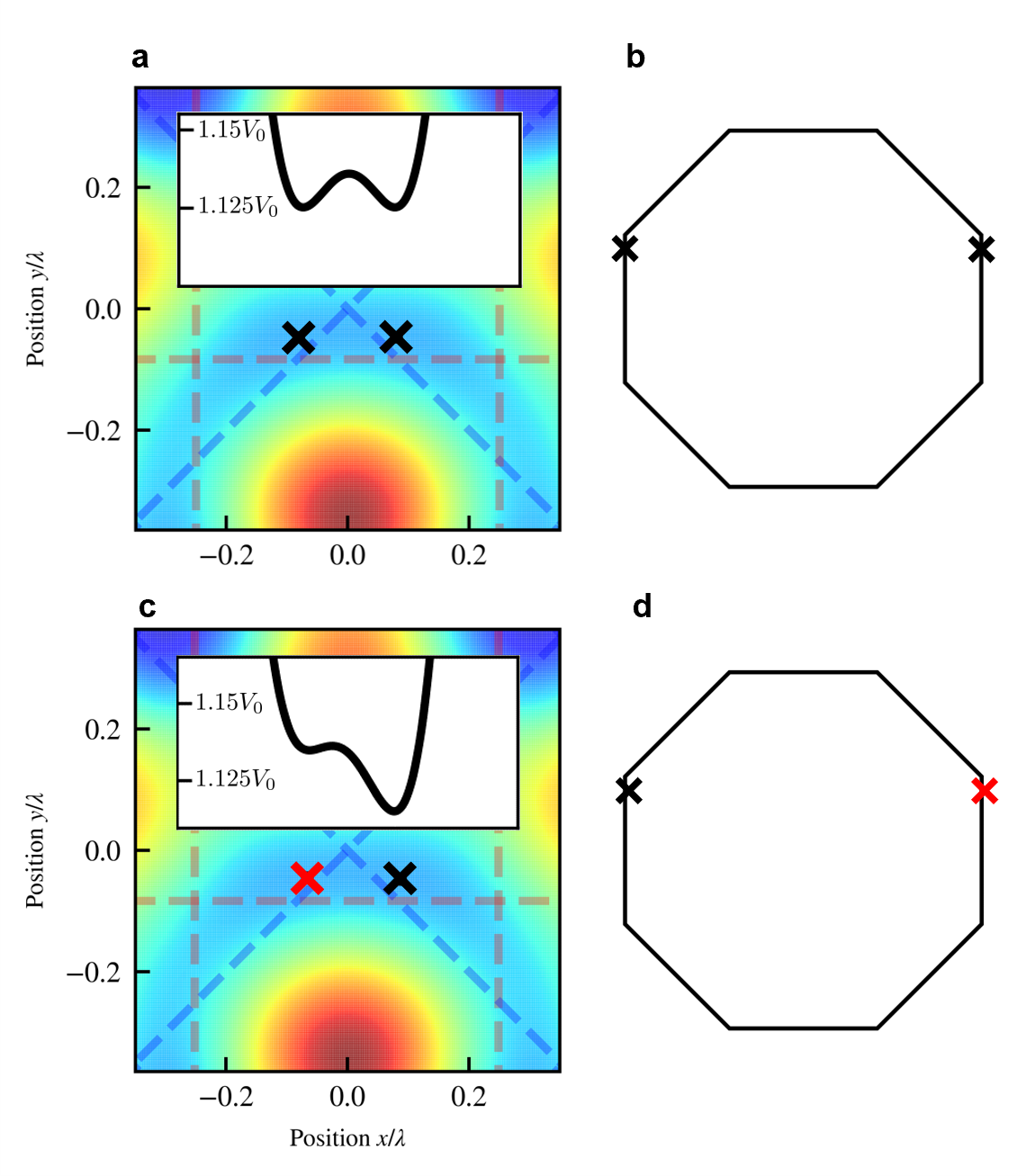}
\caption{Sites (crosses) located close to the edges of the octagon (right) correspond to very shallow double wells in real space (left), which contain only one state in the lowest band. (a, b) shows the case of a perfectly symmetric double well, lying exactly on the edge of the octagon. Both minima (black crosses) are separated by a shallow barrier of height $\approx 0.01 V_0$. (c,d) shows an asymmetric double-well, one of the minima (red cross) lies outside of the octagon. Insets show a 1D cross-section of the double-well potentials. }
  \label{fig:double_well}
\end{figure}

As mentioned in \cref{sec:2D8-foldQC}, the lowest band of the 8QC lattice contains one state per local minimum of the potential, up to some exceptional lattice sites. Indeed, some very shallow local minima exist that do not host a Wannier function in the lowest band, for an example see the red cross in \cref{fig:double_well} c.

Careful inspections shows that in configuration space these minima are always located just outside the edge of the octagon and that they correspond to the higher minima of asymmetric double wells, see the inset on \cref{fig:double_well} c. The other minimum of the double-well is then always located inside the octagon. Numerically diagonalising a patch containing such a double well shows that only the lower state of this double well contributes to the ground band, while the higher state can be found in the excited band. 
The code accounts for the lowest band state hosted in the double-well by generating one Wannier function localised around the minimum lying inside the octagon. By construction, this state will corresponds to the lower eigenstate of the double well.

The minima sitting exactly on the boundaries of the octagon form perfectly symmetric double-wells in real space, see \cref{fig:double_well} a,b. In this configuration, the symmetric combination belongs to the lowest band, while the antisymmetric combination is part of the excited band. Such configurations of lattice sites have also been observed in discrete octagonal quasicrystals obtained from cut-and-project procedures from hypercubic 4 dimensional lattices \citep{jeon_discovery_2022}. As these constitute a set of measure zero in configuration space  they are statistically irrelevant in the thermodynamic limit and hence require no special treatment.

\begin{figure}
  \centering
  \includegraphics[width = \linewidth]{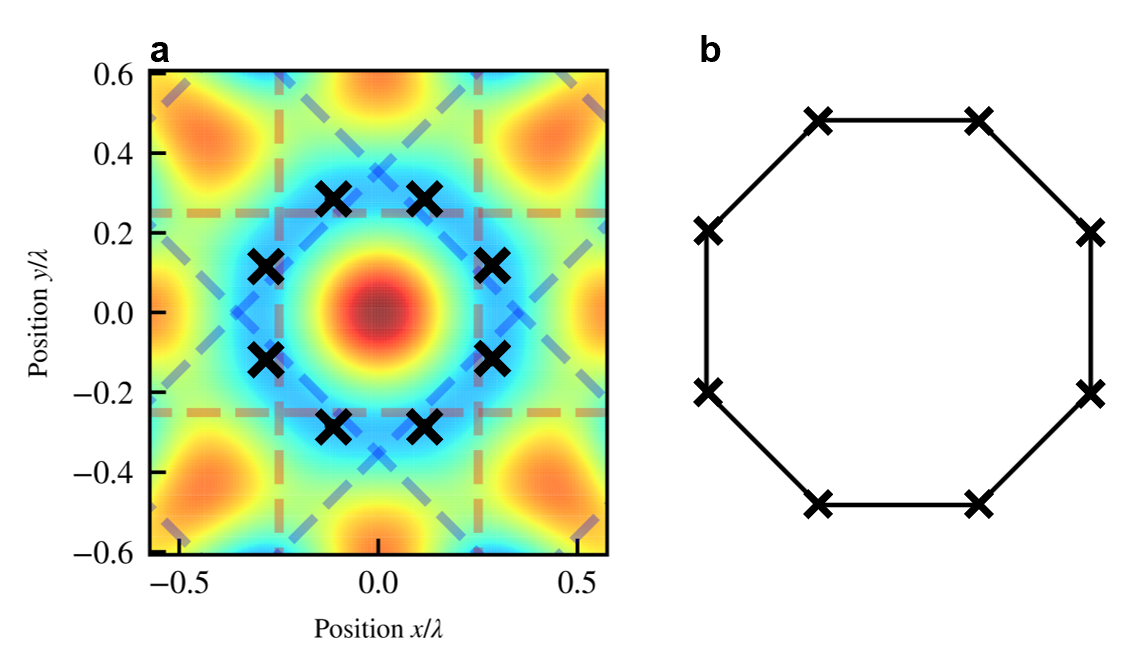}
\caption{Sites (crosses) located on the 8 corners of the octagon (right) form an 8-fold symmetric ring, separated by shallow potential barriers. This constitutes one of the symmetry centers of the 8QC. }
  \label{fig:ring}
\end{figure}

Another specific case arises for the minima located exactly on the eight corners of the octagon. In this case,  the potential forms a perfectly 8-fold symmetric ring containing 8 local minima separated by very weak potential barriers, see \cref{fig:ring}. This configuration is a center of global rotational symmetry of the 8QC, and can hence occur only once. 

\begin{figure}
  \centering
  \includegraphics[width = \linewidth]{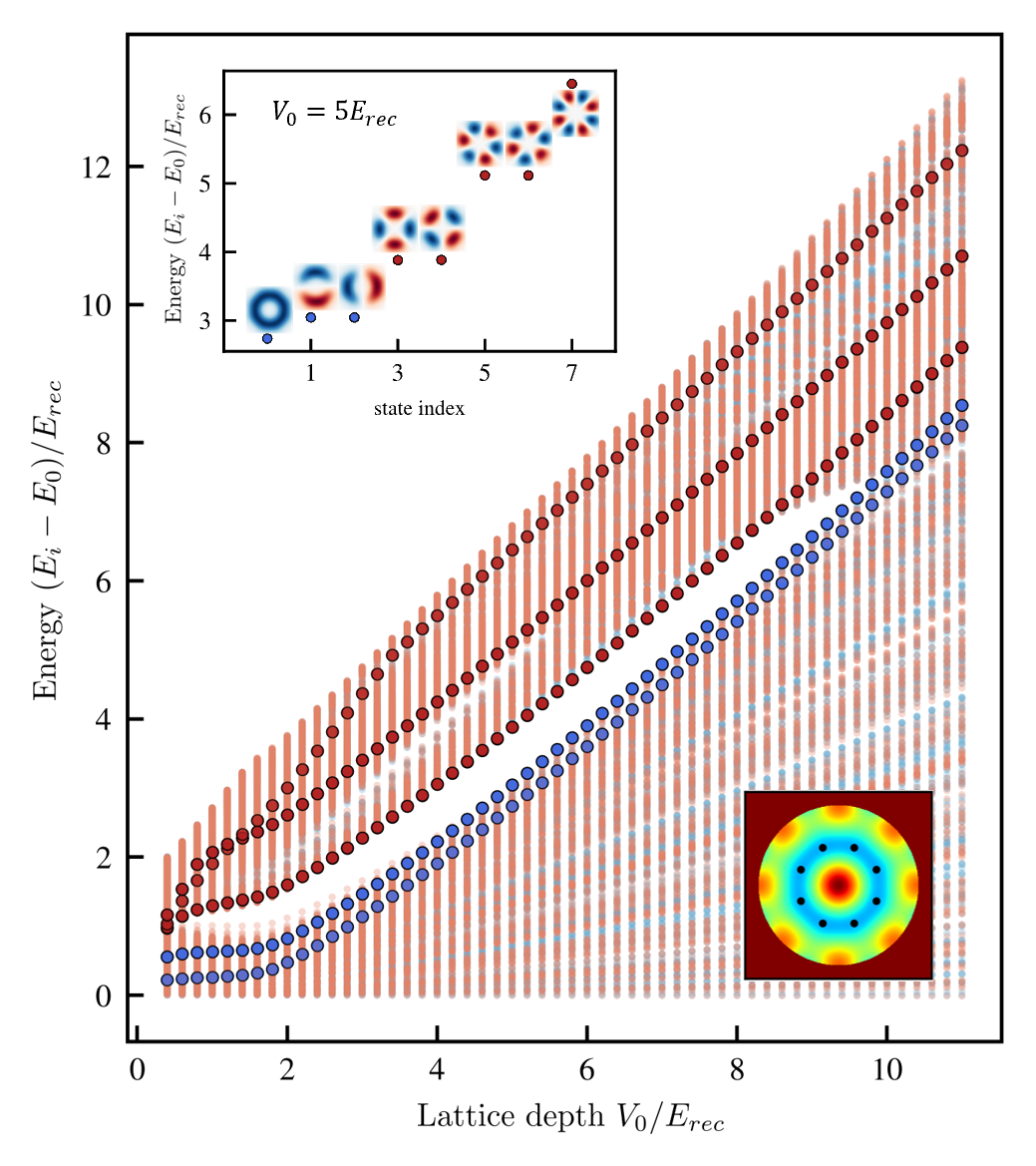}
  \caption{Blue and red circled dots: Energy spectrum of a ring containing 8 minima (inset, lower right) obtained by exact diagonalisation in continuum. Background: bulk energy spectrum of the 8QC from \cref{fig:gapFDS}. The lowest three states of the ring contribute to the lowest band (blue circled dots), while the remaining higher-lying states are located within the higher bands (red circled dots). Upper left inset shows the energy spectrum and the 8 lowest ring eigenstates for $V_0 = 5\,E_{rec}$.}
  \label{fig:ring_spectrum}
\end{figure}

We can obtain an estimate for the energies of the states hosted on this ring using exact diagonalisation of a patch containing the ring. 
\cref{fig:ring_spectrum} shows the energies of theses states as a function of lattice depth, and compares them to the typical energy spectrum of the rest of the lattice. As we see, the ring contains 3 low energy states that are located at the upper limit of the lowest band. The 5 higher-lying states  are instead located within the excited bands. 

Using the classification of 1st-order neighbours developed below, we can see that in situations close to, but not equal to, the 8-fold symmetric ring (or the corner of the octagon), there will naturally be three minima within the octagon while the other 1st-order neighbours lie outside of it. The configuration-space construction hence in all cases automatically selects the right number of minima to reproduce the lowest band.

\section{Neighbours classification and tunneling amplitudes} \label{app:neighbours}
In contrast to for instance the regular square lattice, it is not possible to write down an unambiguous definition for nearest-neighbours in the 8QC based on real-space distances. In configuration space, however, nearest-neighbours and higher order neighbours can be defined rigorously, identically to what is done in the perpendicular space of discrete quasicrystals \cite{oktelStrictlyLocalizedStates2021b}.

We start by rewriting the expressions for $\mathbf{\Phi}_{XY}$ and $\mathbf{\Phi}_D$ (\cref{eq:phases_XY}, \cref{eq:phases_D}) by stating the modulo operation explicitly, and setting all $\phi_i$ to zero for simplicity:

\begin{align}
  \mathbf{\Phi}_{XY} (x,y) &= \left( x - m_1 \ d \right) {\mathbf{e}_x} \nonumber\\
              &+ \left( y - m_2 \ d  \right) {\mathbf{e}_y}
\end{align}
\begin{align}
    \mathbf{\Phi}_{D}(x,y) &= \left( \frac{x+y}{\sqrt{2}} - m_3 \ d \right) {\mathbf{e}_+} \nonumber \\
                & + \left( \frac{x-y}{\sqrt{2}} - m_4 \ d \right) {\mathbf{e}_-}
\end{align}

\begin{figure}
    \centering
    \includegraphics{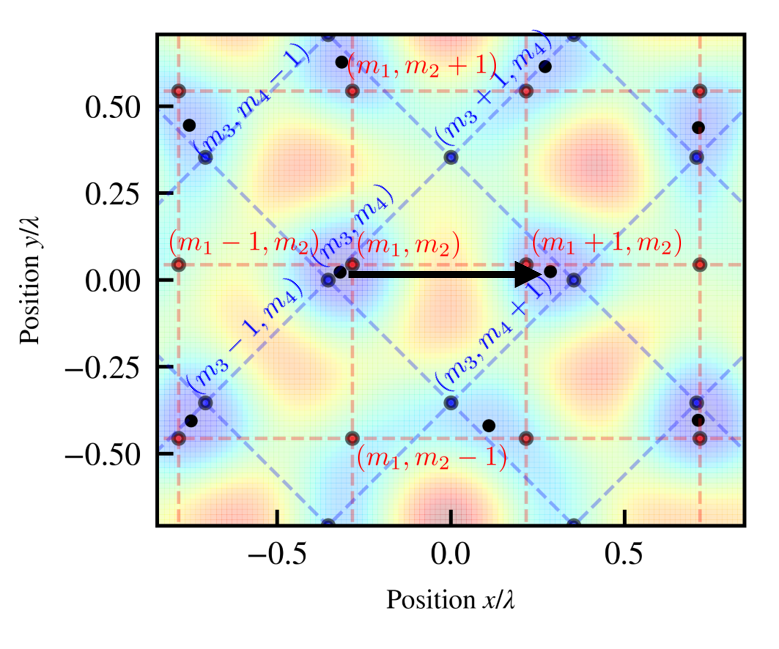}
    \caption{Real-space neighbours (black dots) can be found around adjacent lattice sites of the square lattices (red and blue dots) underlying the 8QC. Red and blue labels indicate the indices of some of the sites adjacent to the $(m_1, m_2)$ site of the XY lattice and $(m_3, m_4)$ site of the D lattice. This illustrates that the nearest neighbour to the right (black arrow) is offset by $\{\Delta m_1,\Delta m_2,\Delta m_3,\Delta m_4\}=\{1,0,1,1\}$ so that 
    $\mathbf{\Phi}' - \mathbf{\Phi}=\mathbf{\tilde{e}}_x$.
    }
    \label{fig:neighbour_construction}
\end{figure}

Here, $m_1,\,m_2\in\mathbb{Z}$ (which are functions of $x$ and $y$) label the lattice sites of the XY lattice, while $m_3$ and $m_4$ label the sites of the D lattice (see \cref{fig:neighbour_construction}), and the $\mathbf{e}_i$ with $i\in\{x,y,+,-\}$ are unit vectors along the four lattice directions in \cref{fig:QClattice}.
In turn, we can re-write the expression for $\mathbf{\Phi} = \mathbf{\Phi}_{XY} - \mathbf{\Phi}_D $.

\begin{align}
  \mathbf{\Phi}(x,y) &= d \left(-m_1 + \frac{m_3+m_4}{\sqrt{2}} \right) \mathbf{e}_x  \nonumber \\
            & + d \left(-m_2 + \frac{m_3-m_4}{\sqrt{2}} 
            \right) \mathbf{e}_y
\end{align}

As an aside, we note that this form also makes it apparent that the octagon is populated densely and uniformly, as the equidistribution theorem \citep{zorziElementaryProofEquidistribution2015} ensures that  the decimal part of sequences of the form $x_n=\alpha n$ with $\alpha$ irrational and $n$ the sequence of natural numbers must be uniformly dense in the interval $[0,1[$.

Let us now consider two neighbouring sites of the 8QC that correspond to the integers $\{m_1,m_2,m_3,m_4\}$, $\{m'_1,m'_2,m'_3,m'_4\}$ (see \cref{fig:neighbour_construction}) and local displacements $\mathbf{\Phi}'$ and $\mathbf{\Phi}$. We can then rewrite the vector connecting $\mathbf{\Phi}'$ to $\mathbf{\Phi}$ as 

\begin{align} \label{eqneighbouroctagon}
    \mathbf{\Phi}' - \mathbf{\Phi} = & d \left(-\Delta m_1 + \frac{\Delta m_3+ \Delta m_4}{\sqrt{2}} \right) \mathbf{e}_x  \nonumber \\
            & + d \left(- \Delta m_2 + \frac{\Delta m_3-\Delta m_4}{\sqrt{2}} 
            \right) \mathbf{e}_y
\end{align}
with $\Delta m_i = m'_i-m_i$. In real-space, nearest neighbours cannot be more than one unit cell of the XY and D lattices away, i.e.,\  $|\Delta m_i|\in\{0,1\}$.

This is clearly visible in \cref{fig:neighbour_construction}. Here, the vector connecting the lattice site defined by $\{m_1,m_2,m_3,m_4\}$ to its right-hand neighbour defined by $\{m_1+1,m_2,m_3+1,m_4+1\}$, amounts to $\mathbf{\Phi}' - \mathbf{\Phi}=- \frac{d}{1+\sqrt{2}}{ \mathbf{e}_x}$. Thanks to eightfold symmetry, the same reasoning can be applied in all 8 directions, leading to:

\begin{equation} \label{eq:etilde} \mathbf{\tilde{e}}_i \in \pm \frac{d}{1+\sqrt{2}} \left\{ \mathbf{e}_x, {\mathbf{e}_y}, {\mathbf{e}_+}, {\mathbf{e}_-} \right\}
\end{equation}.

\begin{figure}
  \centering
  \includegraphics[width = \linewidth]{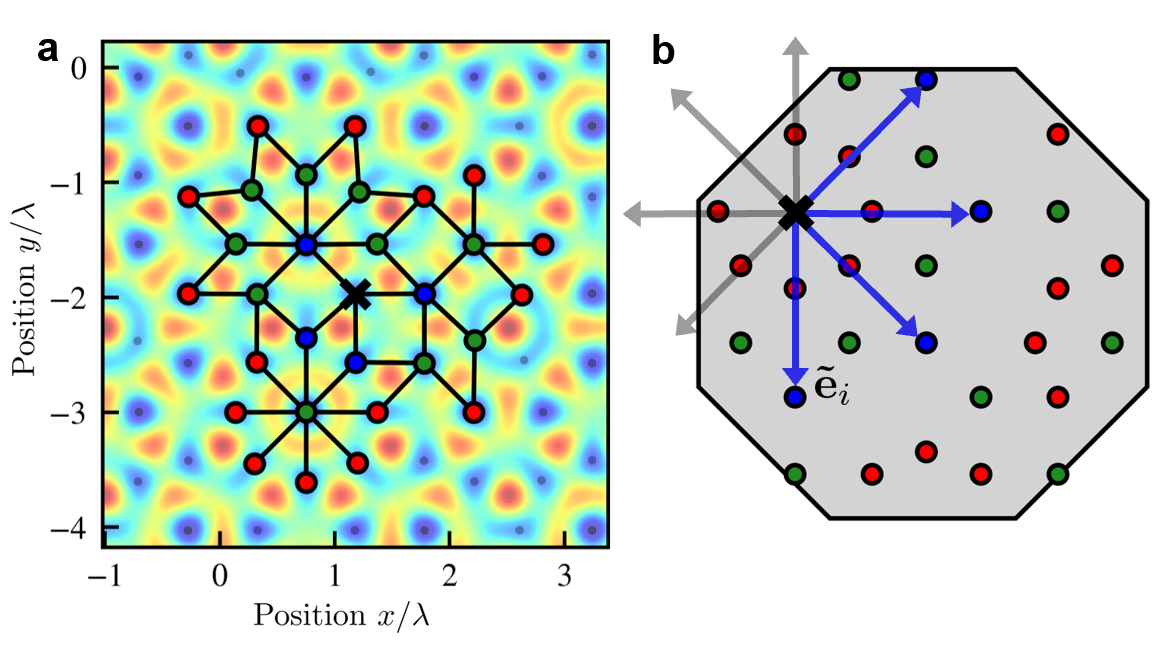}
  \caption{1st, 2nd and 3rd-order neighbours of a given site (black cross) in real (a) and configuration (b) space constructed by $\mathbf{\Phi}' = \mathbf{\Phi}+\sum_i c_i \mathbf{\tilde{e}}_i$ with $c_i\in\mathbb{Z}$ and $\sum{|c_i|}=\{1,2,3\}$. (blue): 1st-order neighbours. (green): 2nd-order neighbours. (red): 3rd-order neighbours. Black lines connect 1-order neighbours.}
  \label{fig:octagon_neighbours}
\end{figure}

\cref{fig:octagon_neighbours} shows that vectors of the form $\mathbf{\Phi}+\mathbf{\tilde{e}}_i$, if they lie within the octagon, do indeed correspond to close neighbours in real space and we accordingly define \textit{first-order} neighbours as sites separated by the vectors $\mathbf{\tilde{e}}_i$ in configuration space. This is identical to the definition of first-order neighbours in the perpendicular space of eightfold discrete quasicrystals \cite{oktelStrictlyLocalizedStates2021b}.

\begin{figure}
  \centering
  \includegraphics[width = \linewidth]{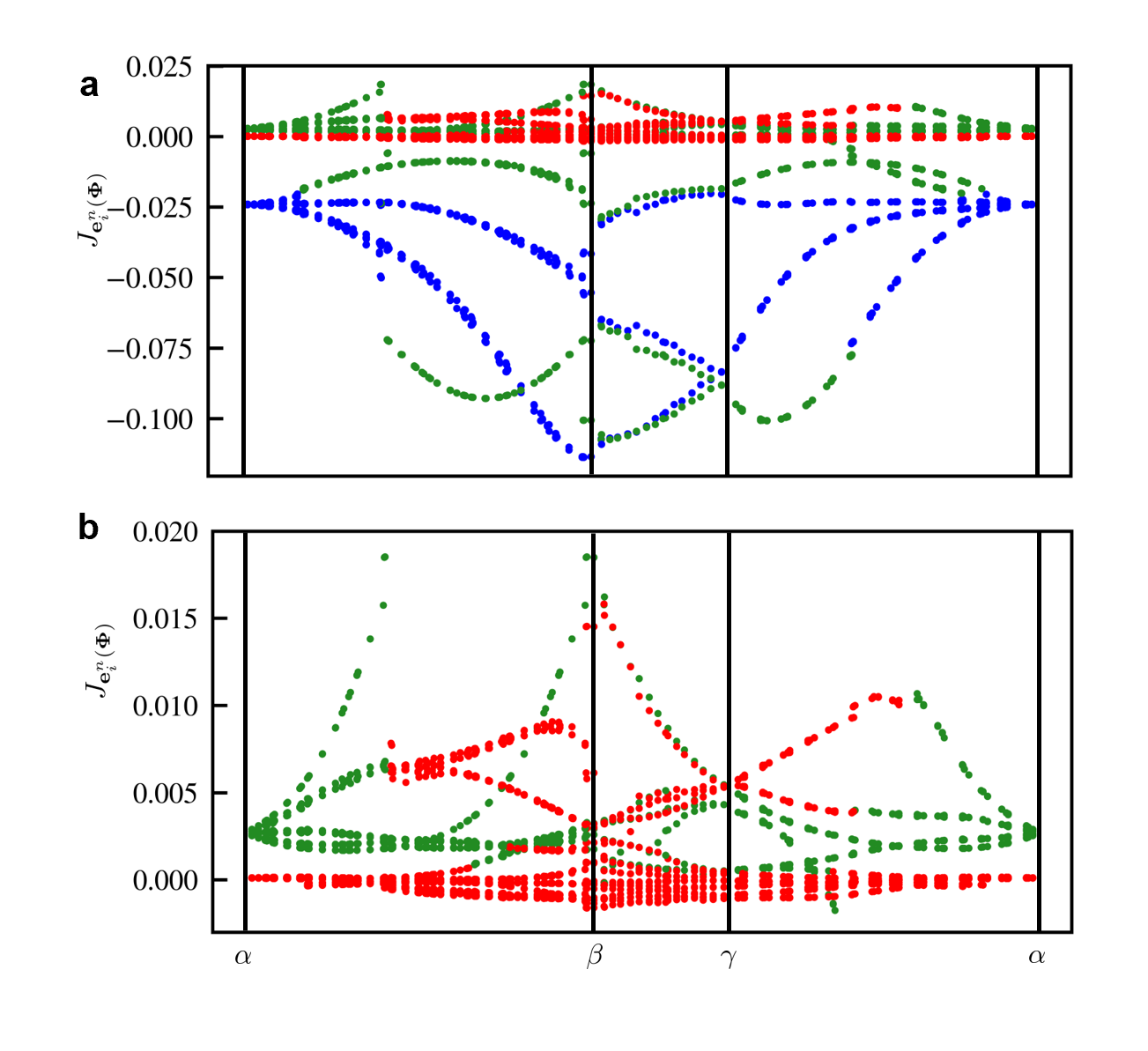}
  \caption{Tunneling amplitudes $J_{\mathbf{e}_k^n}(\mathbf{\Phi}_i) \equiv J(\mathbf{\Phi}, \mathbf{\Phi} + \mathbf{e}_k^n)$ for $V_0 = 2.5\,E_{rec}$ along a cut through configuration space, see \cref{fig:BHoctagon} for definition of $\alpha$, $\beta$ and $\gamma$ points. (blue): 1st-order neighbours. (green): 2nd-order neighbours. (red): 3rd-order neighbours. 1st-order neighbours are always characterised by negative J. (b) Same as (a) but zoomed in on positive tunneling amplitudes.}
  \label{fig:octagon_tunneling}
\end{figure}

In turn, we can define the $n$th-order neighbours of a given site as the set of surrounding sites that lie on the octagon and are connected through the sum of at least $n$ vectors $\mathbf{\tilde{e}}_i$, i.e.,\ $\mathbf{\Phi}' = \mathbf{\Phi}+\sum_i c_i \mathbf{\tilde{e}}_i$ with $c_i\in\mathbb{Z}$ and $\sum{|c_i|}=n$. We refer to the vector connecting n-th order neighbours as $\mathbf{\tilde{e}}_i^n$. \cref{fig:octagon_neighbours} shows an example of 1st, 2nd and 3rd-order neighbours. Using this definition, 1st-order neighbours sit on the edges of the square and rhombuses of the corresponding Ammann–Beenker tiling \citep{maceQuantumSimulation2D2016a}, while second-order neighbours are separated via two edges. Pairs of sites along the short diagonals of the rhombuses are therefore 2nd-order neighbours even though they lie close to each other in real space and give rise to significant (negative) tunneling amplitudes. 

\begin{figure}
  \centering
  \includegraphics[width = .7\linewidth]{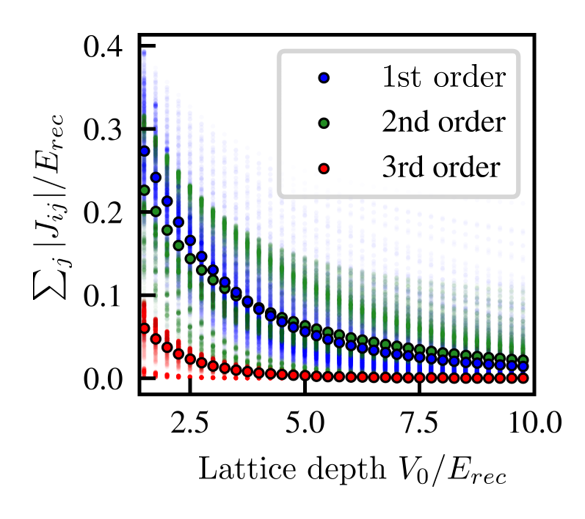}
  \caption{Distributions of 1st, 2nd and 3rd-order total tunneling amplitudes versus lattice depth. Circled dots show the mean values of the distributions. (blue): 1st-order neighbours. (green): 2nd-order neighbours. (red): 3rd-order neighbours.}
  \label{fig:J_orders}
\end{figure}

Similarly to the smooth surfaces formed by the on-site energy $\epsilon(\mathbf{\Phi})$ and the on-site interaction $U(\mathbf{\Phi})$, the tunneling amplitudes connecting two sites $\mathbf{\Phi}, \mathbf{\Phi}'$ can be written as a function $J(\mathbf{\Phi}, \mathbf{\Phi'})$ in configuration space. If we restrict ourselves to $n$th-order neighbours, we know that $\mathbf{\Phi} - \mathbf{\Phi}' = \mathbf{e}_k^n$. Therefore, for each $\mathbf{e}_k^n$ we define a smooth function $J_{\mathbf{e}_k^n}(\mathbf{\Phi}) \equiv J(\mathbf{\Phi}, \mathbf{\Phi} + \mathbf{e}_k^n)$ and  plot a cut through this function in \cref{fig:octagon_tunneling}.

Finally, \cref{fig:J_orders} shows the distributions and mean values for the total tunneling amplitudes connecting 1st, 2nd and 3rd-order neighbours. While 1st and 2nd-order tunneling can have comparable amplitudes, tunneling amplitudes connecting 3rd-order neighbours are significantly weaker for all lattice depths. This also highlights that even though the Ammann–Beenker tiling is bipartite, the 8QC is not.

\section{Relation between configuration and perpendicular spaces of the 8QC}\label{app:perp}

Here we show that the 8QC configuration space introduced in this work directly corresponds to the perpendicular space of discrete octagonal quasicrystals. Both form a densely and uniformly populated octagon, where lattice sites are ordered in terms of their local surroundings. 

Let us first introduce the perpendicular space of discrete octagonal quasicrystals. These quasicrystals can be obtained using a cut-and-project method at an irrational angle of a four dimensional hypercubic lattice \citep{jagannathanEightfoldOpticalQuasicrystal2013a, jeon_discovery_2022}. 

Let $\{\mathbf{e}_1, \mathbf{e}_2, \mathbf{e}_3, \mathbf{e}_4 \}$ be a basis of $\mathbb{R}^4$, and define the hypercubic lattice as the set of their integer combinations.

We then project the 4D hypercubic lattice into two orthogonal subspaces: the "physical space" and "perpendicular" space, using the projection maps $\pi$ and $\pi^\perp$ respectively. These are defined as:  

\begin{equation}
    \mathbf{\pi} = \begin{pmatrix}
    1 &  0 & \frac{1}{\sqrt{2}} & \frac{1}{\sqrt{2}} \\
    0 & 1 & \frac{-1}{\sqrt{2}} & \frac{1}{\sqrt{2}}\\
    \end{pmatrix}
\end{equation}

\begin{equation} \label{eqperp}
    \mathbf{\pi}^\perp = \begin{pmatrix}
    -1 & 0  & \frac{1}{\sqrt{2}} & \frac{1}{\sqrt{2}} \\
    0 & 1 & \frac{1}{\sqrt{2}} & \frac{-1}{\sqrt{2}} \\
    \end{pmatrix}
\end{equation}

The 2D quasicrystalline lattice can then be obtained as the set of physical space positions of the hypercubic lattice sites whose perpendicular space image lies within a certain "acceptance window". A common choice for this window is to set it equal to the perpendicular space image of the hypercubic Wigner-Seitz cell.

Let us now turn the the configuration space of the 8QC. The optical potential can be obtained as an irrational cut of a 4 dimensional hypercubic optical potential (where we fixed all phases $\phi_i$ to $0$ for simplicity)

\begin{equation}
    V_{4D} (x_1,x_2,x_3,x_4) = V_0 \sum_{i=1}^4 \sin^2(\frac{2\pi}{\lambda} x_i)  
\end{equation}

by setting $x_3 = \frac{x_1+x_2}{\sqrt{2}}$ and $x_3 = \frac{x_1-x_2}{\sqrt{2}}$.

In turn, we can rewrite the configuration-space coordinates in 4 dimensions: 

\begin{equation}
    \mathbf{\Phi} = \mathbf{\Phi}_{XY} - \mathbf{\Phi}_D
\end{equation}

with

\begin{equation}
    \mathbf{\Phi}_{XY} = \begin{pmatrix}
    x_1 \; \mathrm{mod}\; d \\
    x_2 \;\mathrm{mod}\; d
    \end{pmatrix} = \begin{pmatrix}
    \tilde{x}_1 \\
    \tilde{x}_2 
    \end{pmatrix}
\end{equation}
\begin{equation}
    \mathbf{\Phi}_D= \frac{1}{\sqrt{2}} \begin{pmatrix}
    x_3 \; \mathrm{mod}\; d + x_4 \; \mathrm{mod}\; d\\
    x_3 \; \mathrm{mod}\; d - x_4 \; \mathrm{mod}\; d
    \end{pmatrix} = \frac{1}{\sqrt{2}} \begin{pmatrix}
    \tilde{x}_3+ \tilde{x}_4\\
    \tilde{x}_3 - \tilde{x}_4
    \end{pmatrix}
\end{equation}

This directly leads to 
\begin{equation}
    \mathbf{\Phi} = \begin{pmatrix}
    \tilde{x}_1 - \frac{\tilde{x}_3+\tilde{x}_4}{\sqrt{2}} \\
    \tilde{x}_2 - \frac{\tilde{x}_3-\tilde{x}_4}{\sqrt{2}} \\
    \end{pmatrix} = 
    \begin{pmatrix}
    1 & 0 & \frac{-1}{\sqrt{2}} & \frac{-1}{\sqrt{2}} \\
    0 & 1 & \frac{-1}{\sqrt{2}} & \frac{1}{\sqrt{2}} \\
    \end{pmatrix}
    \begin{pmatrix}
    \tilde{x}_1 \\
        \tilde{x}_2 \\
    \tilde{x}_3 \\
    \tilde{x}_4 \\
    \end{pmatrix},
\end{equation}
which shows that, up to a sign change, the 8QC configuration space is identical to the perpendicular space projection map $\pi^\perp$ \cref{eqperp}.

\bibliography{apssamp}

\end{document}